\documentclass[useAMS,usenatbib]{mn2e}
\usepackage{graphicx}
\usepackage{pslatex}
\usepackage{natbib}
\usepackage{txfonts}

\newcommand{\vv}[1]{\bmath{#1}}
\newcommand{\mtr}[1]{\mathbfss{#1}}

\begin{document}
\title[Stress field and spin relaxation for ellipsoids]{Stress field and spin axis relaxation for inelastic triaxial ellipsoids}
\author[S. Breiter, A. Ro\.{z}ek and D. Vokrouhlick\'{y}]{S. Breiter$^{1}$\thanks{E-mail:
breiter@amu.edu.pl}, A. Ro\.{z}ek$^{1}$\thanks{E-mail:
 a.rozek@almukantarat.pl} and D. Vokrouhlick\'{y}$^{2}$\thanks{E-mail: vokrouhl@cesnet.cz} \\
$^{1}$Astronomical Observatory Institute, Faculty of Physics, Adam Mickiewicz University, Sloneczna 36, PL60-286 Pozna\'{n}, Poland\\
$^{2}$Institute of Astronomy, Charles University, V Hole\v{s}ovi\v{c}k\'{a}ch 2, 18000 Prague 8, Czech Republic}

\date{}

\pagerange{\pageref{firstpage}--\pageref{lastpage}} \pubyear{2012}

\maketitle

\begin{abstract}
 A compact formula for the stress tensor inside a self-gravitating, triaxial ellipsoid in an arbitrary
 rotation state is given. It contains no singularity in the incompressible medium limit.
 The stress tensor and the quality factor model are used to derive a solution for the energy dissipation
 resulting in the damping (short axis mode) or excitation (long axis) of wobbling.
 In the limit of an ellipsoid of revolution, we compare our solution with earlier ones and show that, with appropriate corrections,
 the differences in damping times estimates are much smaller than it has been claimed.\\
 This version implements corrections of misprints found in the MNRAS published text.
\end{abstract}

\label{firstpage}

\noindent
\textbf{Notice:} The version printed as an article in \emph{Monthly Notices of the  Royal Astronomical Society}
\textbf{427}, 755-769 contains a number of errors:
\begin{itemize}
  \item the sum in Equation (B5) starts from $j=1$ instead of $j=0$,
  \item Figure 1 top (i.e. $T_1(h)$ plot) was traced using an incorrect multiplier -- instead of the correct formula (102)
  with the numerator $1+h_1^2$,
  we used one with $(1+h_1)^2$; this also affects some numbers in the paragraph below Eq. (102),
  \item the set of values $(0.7,1,210)$ in the caption of Fig. 4 should be $(1,0.7,210)$,
  \item the numbers provided in the captions of Fig. 3 and Fig. 4 are actually $(h_1,h_2,1/w)$ instead
  of $(h_1,h_2,w)$ as printed.
\end{itemize}
In the present`postprint' we have fixed these errors.

\section{Introduction}

Most asteroids rotate in the principal, shortest axis mode: their spin axes practically coincide with the directions of
the maximum moment of inertia. Only 45 out of almost 5500 entries of the LCDB light curve database \citep[March 2012 version]{Warner:09}
refer to objects that are possible non-principal axis (NPA) rotators, also known as `tumblers' or wobbling objects. With one exception of 253 Mathilde,
tumblers are rather small, with estimated diameters below 20~km, but even in this size range they belong to
a minority among about 2000 objects of this size with known rotation periods.

Attitude dynamics of asteroids is shaped mainly by gravitational torques (exerted either systematically by the Sun and giant planets, or sporadically during
encounters with other bodies), collisions, optical and thermal radiation recoil torques, i.e. the Yarkovsky-O'Keefe-Radzievskii-Paddack (YORP) effect, and -- last but not least
-- by energy dissipation due to inelastic deformations. As far as NPA rotation is concerned,
collisions and close approaches trigger tumbling \citep{SOWAH:2000,PBW:02}. Small fragments created from collisions of larger
objects are also expected to start their lives in a NPA rotation state.
The YORP effect also excites wobbling \citep{Rub:2000,VBNB:07,BRV:2011}, whereas -- save for possible resonances -- distant bodies gravitation torques are neutral in this respect.
Thus, even accounting for observational selection effects mentioned by \citet{Pravec:05}, the dissipative damping seems to override other effects in most of cases.

The mechanism of wobble  damping was first identified by \citet{Prend:58}. In NPA rotation, the centrifugal acceleration oscillates periodically, deforming each body fragment.
The deformation is not perfectly elastic, so some fraction of fluctuating strain-stress energy is dissipated during each precession period and converted into heat. Draining the elastic energy
affects the kinetic energy of rotation which also decreases. Thus the rotation axis is driven towards the minimum energy state -- rotation around the principal axis of maximum inertia.
The angular momentum, however, is not affected by the energy dissipation, as far as we ignore thermal radiation and consider the body as an isolated system. Prendergast provided
a general form of energy dissipation rate equation for an oblate spheroid\footnote{In this paper we use the word `spheroid' for an arbitrary ellipsoid of revolution.}
based upon the solution of 3D elasticity equations and the assumption that a constant fraction of the oscillating part
of elastic energy is dissipated at each precession period. The latter assumption defines the now commonly adopted `$Q$-model'.

\citet{BuSaf:73} built upon the general idea of Prendergast using combination of a spheroidal shape for rotation and a bent slender beam approximation for elastic energy.
Their simple estimate of spin axis alignment time is still in use -- sometimes in the version provided by \citet{Harris:94}. However, some scepticism
towards it has been brought by observations of asteroids that do rotate around the principal axis in spite of having Burns-Safronov damping time estimate longer than the
age of the Solar System. Meanwhile, the problem migrated to geophysics (e.g. Chandler wobble damping), rotation dynamics of comets and interstellar dust grains physics.
The last branch, stemming from \citet{Purc:1979}, was finally brought back to the dynamics of comets and asteroids with the sequence of papers by Efroimsky and Lazarian
\citep{LazEf:1999,Efroimsky:2000,EfLaz:2000,Efroimsky:2001,Efroimsky:2002}. Their main point of novelty is the attempt to discuss a triaxial object,
represented by a rectangular prism (brick), by solving the complete, quasi-static stress tensor equation \citep{Efroimsky:2000}.
Later on, \citet{MoMo:03} issued the damping model for a spheroid using the same starting point as
\citet{Prend:58}, i.e. solving equations for displacements. The work of \citet{SBH:05} not only provides the solution for a spheroid with two different ways
of estimating the peak elastic energy required for the $Q$-model, but it also offers a long discussion of shortcomings and problems
related with earlier papers mentioned in this paragraph.

Trying to combine the YORP effect with a damping mechanism, we first intended to use the spheroid based  model of \citet{SBH:05}
for arbitrary shape asteroids. This approach, mentioned in \citet{VBNB:07}, became less appealing after a closer inspection,
because of substantial difference in the dynamics of bodies with and without axial symmetry.
On the other hand, the solution of \citet{Efroimsky:2000}, albeit referring to  a triaxial shape, exhibits a number of drawbacks:
\begin{enumerate}
\item As a consequence of using a non-smooth, brick-shaped object, the solution of stress equations is inexact, with unknown error bounds.
\item Compatibility conditions are not fulfilled, i.e. there is no displacements field that might produce the strain tensor found by Efroimsky \citep{SBH:05}.
\item Rotation dynamics is treated by approximate formulae valid only in the neighborhood of principal axis.
\end{enumerate}
Later on, \citet{Efroimsky:2001} suggested Fourier series involving the Jacobi nome as a remedy
for the last item, but none of subsequent works has implemented this guideline.
In these circumstances, we have decided to resume the problem at the point where Efroimsky has abandoned it,
not only using Fourier series to resolve the last problem, but also applying
the triaxial ellipsoid shape which resolves the first two objections as well.  From this point of view,
the present work combines a stress solution in the style of \citet{SBH:05} with energy dissipation treatment
in the spirit of \citet{Efroimsky:2001}.

In Section~\ref{Sec:2} we first formulate the problem of determining the stress tensor and enumerate the assumptions, hoping to
help a reader less familiar with elasticity problems. Basic facts are recalled according the textbooks of \citet{LaLi:7}, \citet{Saad:05}, and \citet{Wilma}.
Two independent methods (displacements approach and stress approach) are used to derive and cross-check the final expressions of the stress tensor.

The $Q$-model of energy dissipation is introduced in Section~\ref{Sec:3} and applied in Section~\ref{Sec:4} to derive an energy dissipation rate formula.
Section~\ref{Sec:5} presents wobble damping time equations based upon the results of Section~\ref{Sec:4} and some exemplary results.
In Section~\ref{Sec:6} we present the reduction to a specific case of a spheroid where a comparison of
our solution with those reported in earlier works is possible.
We use this opportunity to resolve controversies concerning drastically different energy dissipation rates in various models.

\section{Linear elastic model of rotating deformable ellipsoid}

\label{Sec:2}

\subsection{Basic terms: strain, stress, and body forces}

In an arbitrary reference frame, we consider a body as a dense union of material points. Let the set of position vectors $\vv{r}$ define a reference
state (configuration); if the points, for any reason, move with respect to the reference state, their
position vectors will be incremented by displacement vectors $\vv{u}(\vv{r},t)$, dependent on time $t$ as well as on position $\vv{r}$,
creating a new state with
\begin{equation}\label{rprime}
    \vv{r}'(t) = \vv{r} + \vv{u}(\vv{r},t).
\end{equation}
The notion of displacements is too general, because it may include rigid body motion -- translation and rotation of the entire body.
The rigid body motions are discarded by the introduction of the strain $\mtr{e}$ -- the dimensionless, tensor quantity describing deformations
of an infinitesimal volume element in terms of displacements gradient.

\underline{Assumption 1:} The gradient of displacements is small and its square can be neglected.

Under the above assumption, strain tensor $\mtr{e}$  is symmetric by its definition
\begin{equation}\label{epsdef}
\mtr{e} = \frac{1}{2} \left[ \nabla \vv{u} +  \left(\nabla \vv{u} \right)^\mathrm{T}\right],
\end{equation}
where the derivatives are taken with respect to the components of $\vv{r}$.

Two kinds of forces have to be considered in a continuous medium: volumetric forces and surface forces, known also as body forces and tractions,
respectively. Body forces represent `external' force field. In our case they include self-gravitation and forces of inertia.
They are defined as a vector field and specified in terms of their volumetric density $\vv{b}(\vv{r},t)$ -- force divided by the mass of the
volume element, so the total volumetric force
$\vv{F}$ acting on a body with density $\rho(\vv{r})$ is the result of volume integral
\begin{equation}\label{totF}
    \vv{F} =   \int_V \rho \vv{b}\, \mathrm{d}V.
\end{equation}
Surface traction $\vv{t}^{\vv{n}}$ is a vector of the force
acting on an infinitesimal oriented surface,
divided by the area. It describes interactions between adjacent volume elements or forces applied directly on the boundary.
The surface is defined by its unit normal vector $\vv{n}$, and may lie either on a boundary or inside the body.
In order to describe traction at each possible direction of a plane passing through the point $\vv{r}$, each component vector of
$\vv{t}^{\vv{n}}$ in a given basis is projected on each component vector of $\vv{n}$, creating the Cauchy stress tensor $\mtr{T}$.
Thus, traction on a surface defined by $\vv{n}$ can be obtained from $\mtr{T}$  as (unless explicitly stated, repeated index summation
assumed in all formulae)
\begin{equation}\label{trac}
    t_i^{\vv{n}} =  T_{ji}\,n_j.
\end{equation}
The units of stress tensor components are those of force per area.
A good illustration of the two forces nature is a glass of water: body force density is constant throughout the volume (homogenous gravitational field),
whereas tractions define a hydrostatic pressure -- vanishing on the top surface, reaching maximum at the bottom and depending, as a vector,
on the direction of the surface element.

\underline{Assumption 2}: The deformable body forms an isolated system without internal heat sources.

The consequences of this assumption are numerous. First of all, we can use the linear momentum conservation principle in the form of
Cauchy equation, linking stress tensor, body forces and the acceleration of mass particles in an inertial reference frame
\begin{equation}\label{dyn}
    \nabla \cdot \mtr{T} + \rho \vv{b} = \rho \left( \ddot{\vv{r}} + \ddot{\vv{u}} \right).
\end{equation}
If the reference configuration is not fixed in space and still we want it to define the reference frame
for Cauchy equation, then $\ddot{\vv{r}}$ should be transferred to the body forces $\vv{b}$ (subtracted)
as the density of forces of inertia.
In case of rotation, some Corolis type terms involving $\dot{\vv{u}}$ may also appear
in the right-hand side \citep{Tokis:74}.

\underline{Assumption 3}: Quasistatic approximation.

Quasistatic approximation results from setting $\ddot{\vv{u}}=0$ (and any Coriolis type $\dot{\vv{u}}$)
in the right hand sides of (\ref{dyn}). This simplification was generally adopted since \citet{Prend:58}
and we proceed similarly in the present work. Having included forces of inertia in $\vv{b}$, so that $\ddot{\vv{r}}=\vv{0}$,
we solve a static equilibrium equation for $\mtr{T}$
\begin{equation}\label{qss}
    \nabla \cdot \mtr{T} = - \rho \vv{b}(\vv{r},t),
\end{equation}
although body forces can be time dependent. The validity of equation~(\ref{qss}) can be justified if the solution of the original
Cauchy equation is a sum of `free' acoustic waves of high frequency and forced vibrations whose frequencies -- presumably much lower --  come from $\vv{b}$.
Free vibrations can be neglected from two points of view: either we consider them to have zero amplitudes at some initial moment and then a slow, adiabatic forcing
will not excite them considerably, or -- looking forward to the introduction of some dissipation mechanism -- the high frequency terms will be quickly damped.
In both cases the stationary regime oscillations derived from (\ref{qss}) will have amplitudes that differ
from the actual ones by a small quantity comparable with the ratio of free to forced vibrations periods.

\underline{Assumption 4:} Traction-free surface.

Boundary conditions for the stress tensor will be specified as homogenous Neumann conditions on the surface of a deformable body
\begin{equation}\label{bound}
   \mtr{T}\,\vv{n} = \vv{0}.
\end{equation}
The uniqueness of $\mtr{T}$ as a solution of this boundary problem is guaranteed by general theorems \citep{Saad:05},
but it is not seen immediately from three scalar equations of (\ref{bound}), even if we add an additional property: according to Assumption 2,
the angular momentum is conserved, so the stress tensor should be symmetric,
i.e. $\mtr{T} = \mtr{T}^\mathrm{T}$. At this point, \citet{MoMo:03} felt free to postulate $\mtr{T}=\mtr{0}$, which was harshly criticized
by \citet{SBH:05}.

\underline{Assumption 5:} Hookean constitutive relations.

Let us assume that an asteroid is made of a linear, isotropic elastic material with adiabatic Lam\'{e} shear modulus $\mu$ and
the Poisson ratio $\nu$ describing the compressibility (incompressible materials have the maximum possible $\nu = 0.5$).
With these assumptions, the Hooke's law -- serving as a constitutive equation -- states a linear relation between the strain and stress
\begin{equation}\label{Hooke}
T_{ij} = 2\mu\,\left( \frac{\nu}{1-2\nu}\,\delta_{ij}\,\mathrm{tr}\,{\mtr{e}} +   e_{ij} \right),
\end{equation}
or, conversely,
\begin{equation}\label{Hooki}
e_{ij} = \frac{1}{2\,\mu}\,\left(  T_{ij}- \frac{\nu}{1+\nu}\delta_{ij} \,\mathrm{tr}\,{\mtr{T}} \right),
\end{equation}
where $\delta_{ij}$ is the Kronecker delta.

\subsection{Stress approach vs. displacements approach}

Our first goal is to find the symmetric stress tensor $\mtr{T}$ as a solution of equations (\ref{qss}) with boundary conditions (\ref{bound}).
In the stress approach, the problem is solved directly, by assuming some ansatz on $\mtr{T}$ as a function of $\vv{r}$. But if we accept
Assumption 5, additional conditions have to be imposed. Strain is a mathematically meaningful quantity if there exists a displacement field
that generates it through equation~(\ref{epsdef}). Even without explicit knowledge of $\vv{u}$, this is guaranteed by Saint Venant's compatibility
conditions \citep{Saad:05,Wilma}
\begin{equation}\label{compa}
    \nabla \times \left(\nabla \times  \mtr{e} \right) = \mtr{0},
\end{equation}
providing 6 independent relations between $e_{ij}$. Through the constitutive relations (\ref{Hooke}), compatibility equations provide
the identities that a meaningful stress tensor has to obey in addition to boundary conditions.
In next section we show that (\ref{bound}) and (\ref{compa}) together admit a unique solution for $\mtr{T}$.
The stress approach was applied by \citet{Efroimsky:2000} to the problem of a rotating rectangular prism.
Yet, the postulated form of $\mtr{T}$ satisfied only the Cauchy equation (\ref{qss}); neither boundary conditions,
nor compatibility equations could be satisfied
exactly (the latter were not tested at all) and the level of resulting error remains unknown \citep{SBH:05}.

Another way of solving Cauchy equation is the displacements approach. Using equations~(\ref{epsdef}) and (\ref{Hooke}), we convert the first order
differential equation for stress tensor (\ref{qss}) into a second degree equation for displacements vector field $\vv{u}$, obtaining
the quasistatic Lam\'{e} or Cauchy-Navier equation
\begin{equation}\label{Nav}
      \mu \left[ \frac{\mathrm{tr}(\nabla \vv{u})}{1-2\nu}   +  \nabla^2\vv{u} \right] = - \rho \, \vv{b},
\end{equation}
with boundary conditions
\begin{equation}\label{b:Nav}
 2\nu\,(\nabla \cdot \vv{u} )\,\vv{n} + (1-2\nu)\,\left[ \nabla \vv{u} +  \left(\nabla \vv{u} \right)^\mathrm{T}\right]\,\vv{n} = \vv{0},
\end{equation}
derived from (\ref{bound}). Equations (\ref{Nav}) with only three
Neumann boundary conditions (\ref{b:Nav}) admit a solution $\vv{u}(\vv{r},t)$ which is not unique and an arbitrary rigid motion may be added to displacements.
But since the definition of $\mtr{e}$ involves differentiation, the resulting strain and stress tensors
are uniquely defined regardless of remaining  arbitrary terms. Following \citet{DeNo:87}
we will impose two special conditions: the volume integral of the displacements field should vanish
\begin{equation}\label{com:inv}
    \int_V \vv{u}\,\mathrm{d}V = \vv{0},
\end{equation}
and the moment of displacements should also vanish, i.e.
\begin{equation}\label{mom:inv}
    \int_V \vv{r} \times \vv{u}\,\mathrm{d}V = \vv{0}.
\end{equation}
These  six  conditions aim at suppressing rigid translation and rotation terms in displacements
and allow a unique determination of $\vv{u}$, which is of minor interest for the
stress tensor recovered through (\ref{epsdef}) and (\ref{Hooke}), but gives more insight into the question of the reference configuration choice
and simplify energy and momentum balance discussion.
Up to the ambiguity in the last two conditions, the displacements approach was taken by \citet{Chree:1895},
\citet{DeNo:87}, \citet{MoMo:03}, and \citet{SBH:05}.

We can also observe that restoring the term $\rho \ddot{\vv{u}}$ in equation~(\ref{Nav}), we obtain a quantitative measure of quasistatic approximation error.
The homogenous solution will involve a frequency close to
\begin{equation}
\omega_\mathrm{f} = \sqrt{\frac{\mu}{ \rho a^2}},
\end{equation}
where $a$ is the radius of an object.
If body forces are periodic with frequency $\Omega$ (the precession frequency in our case) then, with $\mu$ of the order of $10~\mathrm{GPa}$,
the ratio $ \Omega/\omega_{\mathrm{f}}$
may be safely considered small.

\subsection{Ellipsoid stress solution}

\subsubsection{Homogenous ellipsoid body forces}

Let the reference configuration be a homogeneous rigid ellipsoid with semiaxes $c \leqslant b \leqslant a$. Its shape will be described by two dimensionless parameters
\begin{equation}\label{semi}
 h_1 = \frac{b}{a}, \qquad h_2 = \frac{c}{b},
\end{equation}
both taking values $0 < h_i \leqslant 1$. In the reference frame whose centre coincides with the centre of mass and the basis vectors $\vv{e}_i$ are directed along the principal axes,
the parametric equation of the interior reads
\begin{equation}\label{par:eli}
    \vv{r} = q\,a\, \left(\sin{\vartheta} \cos{\phi}\, \vv{e}_1 + h_1 \sin{\vartheta} \sin{\phi} \,\vv{e}_2+  h_1\,h_2\,\cos{\vartheta}\,\vv{e}_3  \right),
\end{equation}
where $0 \leqslant q < 1$, $0 \leqslant \vartheta \leqslant \pi$, $0 \leqslant \phi < 2\pi$. The boundary is specified by $q=1$ and the unit normal vector
on the boundary is
\begin{eqnarray}
    \vv{n} & = &  \Phi(h_1,h_2,\vartheta,\phi)\,\left(\sin{\vartheta} \cos{\phi}\, \vv{e}_1  \right. \nonumber \\
    & & \left. + h_1^{-1} \sin{\vartheta} \sin{\phi} \,\vv{e}_2+ h_1^{-1} h^{-1}_2\,\cos{\vartheta}\,\vv{e}_3  \right),
    \label{norv}
\end{eqnarray}
where $\Phi$ is some nonzero function; its explicit definition is not required for the vanishing traction condition (\ref{bound}).
According to the postulate (\ref{com:inv}), the centre of ellipsoid remains the centre of mass even in a deformed state.

Given an arbitrary function $F(\vv{r})$, the volume integration rule for an ellipsoid is
\begin{equation}\label{vol:int}
    \int_V F  \mathrm{d}V = a^3 h_1^2 h_2 \int_0^1 q^2\, \mathrm{d}q \int_0^\pi \sin{\vartheta}\, \mathrm{d}\vartheta \int_0^{2\pi} F\,\mathrm{d}\phi,
\end{equation}
where $F(\vv{r})$ should be expressed in terms of $q,\vartheta,\phi$ according to equation~(\ref{par:eli}).

The body forces acting on a freely rotating ellipsoid include the forces of inertia due to rotation, with force density vector
\begin{equation}\label{b:in}
    \vv{b}_{\mathrm{in}} = \vv{r} \times  \dot{\vv{\omega}}  + \vv{\omega} \times (\vv{r} \times \vv{\omega}),
\end{equation}
where the time derivative of rotation vector $\vv{\omega}$ is given by Euler equations
\begin{equation}\label{Euler}
   \dot{\vv{\omega}} = - \frac{1-h_2^2}{1+h_2^2} \omega_2 \omega_3 \vv{e}_1 + \frac{1-h_1^2 h_2^2}{1+h_1^2 h_2^2} \omega_1 \omega_3 \vv{e}_2 - \frac{1-h_1^2}{1+h_1^2} \omega_1 \omega_2 \vv{e}_3,
\end{equation}
and the moments of inertia $I_i$ for an ellipsoid with mass $m$
\begin{equation}\label{iner}
    \frac{I_1}{m a^2} = \frac{ h_1^2 \, \left(1+h_2^2\right)}{5}, \quad
     \frac{I_2}{m a^2 } = \frac{ 1+h_1^2 h_2^2}{5}, \quad
      \frac{I_3}{m a^2} = \frac{1+h_1^2}{5},
\end{equation}
are substituted.

Gravitation inside the ellipsoid results in body forces density
\begin{equation}\label{grav}
    \vv{b}_{\mathrm{gr}} = - \gamma_1 x\, \vv{e}_1 - \gamma_2 y \,  \vv{e}_2 - \gamma_3 z\,\vv{e}_3,
\end{equation}
with constants
\begin{eqnarray}
    \gamma_1 & = & \frac{\gamma m}{a^3} R_J(1, h_1^2, h_1^2 h_2^2,1), \nonumber \\
    \gamma_2 & = & \frac{\gamma m}{a^3} R_J(1, h_1^2, h_1^2 h_2^2,h_1^2), \label{gravi} \\
    \gamma_3 & = & \frac{\gamma m}{a^3} R_J(1, h_1^2, h_1^2 h_2^2, h_1^2 h_2^2), \nonumber
\end{eqnarray}
expressed in terms of gravitation constant $\gamma$ and Carlson's elliptic integral
\begin{equation}\label{RJ}
    R_J(u,v,w,p) = \frac{3}{2} \int_0^\infty  \frac{\mathrm{d}s}{(p+s) \sqrt{(u+s)(v+s)(w+s)}}.
\end{equation}

The body forces are linear in coordinates of reference configuration, so we write them as
\begin{equation}\label{bfm}
    \vv{b} = \vv{b}_{\mathrm{in}}+\vv{b}_{\mathrm{gr}} = \mtr{B} \vv{r},
\end{equation}
with coordinates independent matrix $\mtr{B}$ having elements
\begin{eqnarray}
  B_{11} &=& \omega_2^2+\omega_3^2 - \gamma_1, \nonumber \\
  B_{22} &=& \omega_3^2+\omega_1^2 - \gamma_2, \nonumber \\
  B_{33} &=& \omega_1^2+\omega_2^2 - \gamma_3, \nonumber\\
  B_{12} &=& -\frac{2 \omega_1 \omega_2}{1 + h_1^2}, \label{Bdef:1} \\
  B_{13} &=& -\frac{2 \omega_1 \omega_3}{1 + h_1^2 h_2^2}, \nonumber \\
  B_{23} &=& -\frac{2 \omega_2 \omega_3}{1 + h_2^2}, \nonumber
\end{eqnarray}
and
\begin{equation}\label{lower}
    B_{21} = h_1^2 B_{12}, \quad B_{31} = h_1^2 h_2^2 B_{13}, \quad B_{32} = h_2^2 B_{23}.
\end{equation}

\subsubsection{More assumptions}

\underline{Assumption 6:} Displacements and their partial derivatives are small quantities of the first order.

This stronger variant of Assumption 1 allows to treat the problem in terms of a first order approximation.
Namely, considering boundary conditions we can impose them on the reference ellipsoid surface, neglecting the displacements that deform it.
It also means we do not restrict reference ellipsoids to a figure of equilibrium type solutions like e.g. Jacobi ellipsoids \citep{Chan:69}.
In the same spirit, we ignore the variations of density due to strain and use a constant, mean $\rho$ whenever it serves to define displacements
or stress -- either explicitly, or indirectly (like in body forces $\vv{b}_{\mathrm{gr}}$).

It turns out that postulates (\ref{com:inv}) and (\ref{mom:inv}) are inherently related with the choice of reference configuration
that satisfies our assumption. In the first approximation, i.e. evaluating volume integrals over the homogeneous reference ellipsoid,
we interpret (\ref{com:inv}) as a postulate that the centre of mass position is not altered by displacements. Similarly, equation~(\ref{mom:inv})
indirectly leads to the statement, that displacements velocities do not contribute to the angular momentum of the system.
Observing that $\dot{\vv{u}}$ will depend on $\dot{B}_{ij}$ in exactly the same form as $\vv{u}$ depends on $B_{ij}$ (the Cauchy-Navier equation is linear and does not involve
time derivative), we find
\begin{equation}\label{mom:inv:1}
    \int_V \rho ( \vv{r} \times \dot{\vv{u}})\,\mathrm{d}V = \vv{0},
\end{equation}
as a consequence of (\ref{mom:inv}), provided $\rho$ is constant and the same bounding surface is used in both integrals, i.e. within the first order approximation.
In other words, the postulate (\ref{mom:inv})
implies that we use Tisserand's mean axes \citep{MuMa:60} as the reference frame and they approximately coincide with the principal axes of the reference ellipsoid.
Such choice has a property of minimizing the displacements and their velocities.

\citet{SBH:05} postulated a pre-stressed state of their reference spheroid and dropped the constant part of the stress due to self-gravitation on the onset of their derivation.
They were not consequent in this point, because they did not do the same with a mean part of centrifugal stress. For typical asteroids both effects
may be comparable. They may even mutually cancel. Thus we do not find the pre-stressed state assumption necessary for asteroids, although it is important for major objects, like the Earth,
where it was originally introduced \citep{Love:34}. On the other hand, its role would be to provide the rationale for the validity of the six assumptions we have already made.

\subsubsection{Displacements approach}

The problem of finding displacements and stress tensor for a freely rotating ellipsoid (without self-gravitation) was solved by \citet{DeNo:87}.
Their impressive solution, apparently worked out without the support of computer algebra, was based upon the displacements approach,
which helped in establishing the influence of displacements field on the moments of inertia.
Earlier results of \citet{Chree:1895} included the self-gravitation, but only the principal axis rotation was considered.
We used the stress tensor of \citet{DeNo:87} as a test of our solution.

The main advantage of using an ellipsoid is the possibility of finding the
displacements field that satisfies equations~(\ref{Nav}), (\ref{b:Nav}) and (\ref{bfm}) as a sum of two homogeneous polynomials
of degrees 1 and 3 i.e. a sum of 13 monomials with dimensionless vector coefficients $\vv{f}$
\begin{eqnarray}
   \vv{u} & = & \vv{f}^{100} x + \vv{f}^{010} y + \vv{f}^{001} z  + a^{-2}\,\left( \vv{f}^{003} z^3  + \vv{f}^{012} y z^2   +  \vv{f}^{021} y^2 z \right. \nonumber \\
   & & + \vv{f}^{030} y^3  + \vv{f}^{102} x z^2  + \vv{f}^{111} x y z +
 \vv{f}^{120} x y^2  + \vv{f}^{201} x^2 z  \nonumber \\
 & & \left.  + \vv{f}^{210} x^2 y  +
 \vv{f}^{300} x^3 \right),
 \label{uh:def}
\end{eqnarray}
which involves 39 arbitrary constants: 9 for the linear part and 30 for the cubic terms. It is instructive to add $B_{21}$, $B_{31}$, and $B_{32}$ as
additional, unspecified parameters, raising the number of unknowns to 42.  Our choice for the form of $\vv{u}$
is straightforward, although not necessarily optimal; for example, \citet{SBH:05} used the spherical harmonics basis
and reported a smaller number of undetermined coefficients  for a spheroid and
in an unpublished solution for the ellipsoid (Sharma, private communication).

Using the `brute force' attack, allowed by the use of an algebraic manipulator,
we formulate 42 independent conditions involving linearly the coefficients of $\vv{u}$ and unspecified elements $B_{ij}$ of matrix $\mtr{B}$ -- the ones with $i \leqslant j$ are parameters,
and those with $i > j$ are unknowns. First, we observe that the centre of mass
condition (\ref{com:inv}) is satisfied identically by (\ref{uh:def}). Actually, it has been used  implicitly  to drop
coordinate independent and quadratic parts of $\vv{u}$.
\begin{enumerate}
\item Equation (\ref{Nav}) generates three equations linear in $x$, $y$, and $z$. This amounts to 9 equations linking each
$B_{ij}$ with 5 coefficients of the cubic part of $\vv{u}$.
\item Postulating equation~(\ref{mom:inv}), we obtain 3 equations -- each linking two coefficients of the linear and six of the cubic part of $\vv{u}$.
\item The remaining 30 equations result from boundary conditions (\ref{b:Nav}). They can be derived either directly, i.e. from the polynomial form
of ellipsoid surface and normal vector \citep{DeNo:87}, or by considering trigonometric polynomials of $\phi$ and $\vartheta$ resulting from
the substitution of equations~(\ref{par:eli}) and (\ref{norv}). In the latter case, we first equate to 0 the coefficients of $\sin{3 \phi}$, $\cos{3\phi}$, $\sin{2 \phi}$ and
$\cos{2 \phi}$ which have a common factor $(\sin \vartheta)^3$ or $(\cos \vartheta)^3$; this provides 12 conditions.
Then, in the coefficients of $\sin{\phi}$ and $\cos{\phi}$,
we separate the terms with $\sin{\vartheta}$ and $\sin{3 \vartheta}$, and set them to 0, obtaining another 12 conditions. Finally, in
the part independent on $\phi$ we separate terms factored by $\cos{\vartheta}$ and $\cos{3 \vartheta}$ resulting in the last 6 conditions.
\end{enumerate}
Solving the Cramer system of 42 linear equations we obtain a unique solution for the 39 coefficients of $\vv{u}$ as linear combinations of $B_{i,j}$ with $i \leqslant j$.
For the remaining three unknowns we recover equation~(\ref{lower}) which becomes the condition of existence of the displacements solution (\ref{uh:def}) for an ellipsoid
-- slightly more general than the condition found by \citet{DeNo:87}.

Unfortunately, the resulting expressions of displacements field $\vv{u}$ are too long to be quoted. Each $B_{ij}$ appearing in $\vv{f}$ has its own multiplier -- a rational function of
$h_1^2$, $h_2^2$ and $\nu$. The only common factor of all $\vv{f}$ vectors is $\rho\,a^2\,\mu^{-1} = \omega_{\mathrm{f}}^{-2}$. This indicates a link between
Assumptions 3 and 6.

\subsubsection{Stress approach}

The structure of displacements field is inherited by the stress and strain tensors. Each $T_{ij}$ and $e_{ij}$
is a linear combination of $B_{ij}$, including zero and second degree monomials of $x$, $y$, and $z$. We focus the discussion on the stress tensor,
because $\mtr{T}$  can be of interest in studying the breakup of spinning asteroids (e.g. \cite{WaSch:2002}), whereas strain components $e_{ij}$
in the elastic material are easily found from $\mtr{T}$ using constitutive relations (\ref{Hooki}).

Introducing
\begin{equation}\label{til}
    \tilde{y} = \frac{y}{h_1}, \qquad \tilde{z} = \frac{z}{h_1 h_2},
\end{equation}
to benefit from some symmetries, we can write the general form of $\mtr{T}$ for the elastic ellipsoid as
\begin{eqnarray}
    \mtr{T} & = &  \rho \, \left( a^2 \mtr{A} - x^2 \mtr{A}^{11} - \tilde{y}^2 \mtr{A}^{22} - \tilde{z}^2 \mtr{A}^{33} \right. \nonumber \\
    & & \left.  - x \tilde{y} \mtr{A}^{12}
     - \tilde{y}\tilde{z} \mtr{A}^{23} - x \tilde{z} \mtr{A}^{13} \right), \label{Tform}
\end{eqnarray}
where $\rho a^2 \mtr{A}$ represents the stress tensor at the origin $x=y=z=0$, and all matrices are symmetric.
Thus the stress definition requires 42 matrix elements to be determined. Their expressions resulting from direct substitution of displacements solution are
unwieldy, so we decided to apply the second possible way of determining $\mtr{T}$ -- the stress approach.
Quasistatic equilibrium condition (\ref{qss}) does not involve $\mtr{A}$ and leads to 9 linear relations between $\mtr{A}^{ij}$ and body force matrix $\mtr{B}$.
Boundary conditions (\ref{bound}) generate 30 relations between the elements of $\mtr{A}$ and $\mtr{A}^{ij}$. However, the resulting set of 39 linear equations
admits solutions only if conditions (\ref{lower}) are satisfied and then its rank drops to 36 allowing a unique solution for all $\mtr{A}^{ij}$
as linear combinations of $\mtr{A}$ and $\mtr{B}$ elements.

`Central stress' $\mtr{A}$ results from compatibility equations (\ref{compa}), forming two independent subsystems of equations:
\begin{eqnarray}
  \frac{\partial^2 e_{11}}{\partial y \partial z} &=& \frac{\partial^2 e_{12}}{\partial x \partial z} - \frac{\partial^2 e_{23}}{\partial x^2}
  + \frac{\partial^2 e_{13}}{\partial x \partial y}, \nonumber \\
  \frac{\partial^2 e_{22}}{\partial x \partial z} &=& \frac{\partial^2 e_{23}}{\partial x \partial y} - \frac{\partial^2 e_{13}}{\partial y^2}
  + \frac{\partial^2 e_{12}}{\partial y \partial z}, \label{compa:1}\\
    \frac{\partial^2 e_{33}}{\partial x \partial y} &=& \frac{\partial^2 e_{13}}{\partial y \partial z} - \frac{\partial^2 e_{12}}{\partial z^2}
  + \frac{\partial^2 e_{23}}{\partial x \partial z}, \nonumber
\end{eqnarray}
and
\begin{eqnarray}
  2 \frac{\partial^2 e_{12}}{\partial x \partial y} &=& \frac{\partial^2 e_{11}}{\partial y^2} + \frac{\partial^2 e_{22}}{\partial x^2}, \nonumber \\
  2 \frac{\partial^2 e_{23}}{\partial y \partial z} &=& \frac{\partial^2 e_{22}}{\partial z^2} + \frac{\partial^2 e_{33}}{\partial y^2}, \label{compa:2}\\
  2 \frac{\partial^2 e_{13}}{\partial x \partial z} &=& \frac{\partial^2 e_{11}}{\partial z^2} + \frac{\partial^2 e_{33}}{\partial x^2}. \nonumber
\end{eqnarray}
Relating $\mtr{e}$ with $\mtr{T}$ by means of (\ref{Hooki}), substituting the general form of stress (\ref{Tform}) and making use of $\mtr{A}^{ij}$ expressed in terms
of $\mtr{A}$ and $\mtr{B}$, we find that subsequent equations of (\ref{compa:1}) directly define $A_{23}$, $A_{13}$, and $A_{12}$, respectively, in terms of $B_{ij}$
with the same subscripts. On the other hand, equations (\ref{compa:2}) form a system of coupled linear equations for $A_{ii}$ with right hand sides depending on $B_{ii}$.
Its solution takes form
\begin{equation}\label{ouch}
   \left( \begin{array}{c} 2 A_{11} \\ 2 h_1^{-2}A_{22} \\ 2 h_{12}^{-2} A_{33} \end{array}\right)
    =   \left( \mtr{1} + \mtr{L}^{-1} \mtr{R} \right)
     \left( \begin{array}{c} B_{11} \\  B_{22} \\ B_{33} \end{array} \right),
\end{equation}
requiring the inverse of a  $ 3 \times 3$ matrix $\mtr{L}$. Although $\mtr{L}$ is not complicated and invertible by elementary means, an explicit formula
for $\mtr{L}^{-1}$ is too long to be explicitly quoted.
The detailed solution for $\mtr{A}^{ij}$ and matrices $\mtr{L}$, $\mtr{R}$ are given in Appendix~\ref{ap:A}.

We have verified that both methods (displacements approach and stress equations with compatibility conditions)
lead to the same final results.
Nevertheless, the reduction of $\mtr{T}$ to the simple form given in Appendix~\ref{ap:A} starting from the displacements solution would be a tedious exercise.

\section{The $Q$ model of dissipation}
\label{Sec:3}

\subsection{Energy of a deformable ellipsoid}

For a rigid ellipsoid in free rotation with angular velocity $\vv{\omega}$, the only part of energy that matters is kinetic energy
\begin{equation}\label{K0}
    K_0 = \frac{1}{2} \int_V \rho \, (\vv{\omega} \times \vv{r})^2\,\mathrm{d}V =  \frac{1}{2}  \omega_i I_{ij} \omega_j,
\end{equation}
conserved during the motion. In the principal axes system, the tensor of inertia $\mtr{I}$ is diagonal, i.e $I_{ij} = \delta_{ij}I_i$, given in equation~(\ref{iner}).
In the absence of external torques, the angular momentum
\begin{equation}\label{angmom}
    \vv{H} = \int_V \rho \, \vv{r} \times (\vv{\omega} \times \vv{r})\,\mathrm{d}V = I_{ij} \omega_j \vv{e}_i,
\end{equation}
is also conserved in an inertial frame, i.e. provided we account for the rotation of body-fixed basis vectors $\vv{e}_i$. Only the norm $H=||\vv{H}||$ is constant in the body frame.

In a deformable body, the velocity of displacements $\dot{\vv{u}}$ adds up to the total kinetic energy $K$ and angular momentum. Moreover, the results of volume integration
are affected by the fluctuations of density and by the bounding surface deformations. However, under Assumption 2, as long as the deformations are elastic and the centre of mass is not altered
by displacements, the angular momentum remains constant in the fixed frame. Total kinetic energy may vary, but the sum $K+U$, where $U$ is the potential energy of deformation, remains constant.
A rigorous discussion of energy exchange between $K$ and $U$ can be found in  \citet{MuMa:60} or \citet{Lambeck:80}.
For the present discussion, we approximate the kinetic energy as $K \approx K_0$ and consider
potential energy $U \approx U_{\mathrm{e}}$, storing the work of both body forces and tractions,  to be the elastic energy
\begin{equation}\label{elase}
    U_{\mathrm{e}}  = \int_V \epsilon \,\mathrm{d}V  \geqslant 0,
\end{equation}
with the energy density $\epsilon$ defined  as
\begin{equation}\label{eden}
    \epsilon =  \frac{1}{2}  e_{ij} T_{ij},
\end{equation}
which is evaluated, according to the first order approximation, with the volume integral taken over the reference ellipsoid.
Under the Assumption 5, we can also express $\epsilon$ in terms of the stress tensor alone, obtaining \citep{Efroimsky:2000}
\begin{equation}\label{edens}
    \epsilon = \frac{1}{4\mu}\,\left( - \frac{\nu \,( \mathrm{tr}\,\mtr{T} )^2 }{1+\nu} + T_{11}^2+T_{22}^2+T_{33}^2 + 2 \,\left( T^2_{12}+ T^2_{23}+T^2_{13} \right) \right).
\end{equation}

Inelasticity disturbs the ideal picture of Hookean oscillations. Different rheological models have been
constructed in hope to adjust constitutive relations between stress and strain to the reality of the material world.
Extended constitutive relations are formulated either as differential equations, involving $\dot{\mtr{T}}$ and/or $\dot{\mtr{e}}$,
or in terms of integrals of creep and relaxation functions. This leads to the occurrence of a time lag between
forced oscillations of strain and stress. In this respect, the situation becomes similar to a periodically driven harmonic oscillator
with damping: due to a lag between velocity and position, the time derivative of potential energy integrated over a forcing cycle does not vanish
and generates a power deficit credited by the external forcing in order to sustain stationary oscillations. In the case of our study,
the power supply comes from the kinetic energy $K_0$ and is not unlimited. Moreover, a coupling exists between the power supply and demand,
because the amplitude of stationary vibrations depends on the excess of  kinetic energy  over the ground state of $\frac{1}{2} I_3 \omega^2$.

\subsection{Quality factor principle}

Introducing the quality factor $Q$ as an empirical parameter, one may,  in principle, discuss the energy dissipation in vibrating inelastic materials
without any explicit knowledge of their constitutive relations. In practice, however, the problem of unknown rheology leaks into the question of a proper
definition of $Q$ and of its dependence on driving frequency, temperature, etc. \citep{EfroWil:09}.
In the limit, a perfect definition of $Q$ is probably not easier than defining an adequate rheological model and solving the related inelastic
oscillations problem.

\citet{OCB:78} tried to put some order into a growing number of different $Q$ definitions.
They warned about `confusion between some ill-defined
$Q$ of a process' and `an intrinsic $Q$ of the material'. Their generally acclaimed definition
of quality factor as the ratio of real and imaginary parts of compliance leads
`for a large class of viscoelastic materials' to the common rephrasing as
\begin{equation}\label{Qdef:1}
    Q  = 2 \pi \frac{(2 E_\mathrm{av})}{\Delta E},
\end{equation}
where $\Delta E$ is the energy lost during one period of a harmonic (pure sine) loading, and $E_\mathrm{av}$
is the `average stored energy' during the loading cycle \citep{OCB:78}. The difficult point of this apparently
simple definition is how to interpret $E_\mathrm{av}$ in some particular problem, even if we use the strain
energy (\ref{elase},\ref{eden}) for this purpose.

\citet{SBH:05} complained that `definitions of measures of energy fluctuations corresponding to the type of loadings
encountered with tumbling bodies are not readily available', meaning the presence of a constant term in body
forces that are not purely periodic.  Although their main solution followed the usual habit of dropping
the constant part from strain and stress when plugging $U_\mathrm{e}$ into the definition (\ref{Qdef:1}),
they express serious doubts and find `no easily identifiable reason' for it,
except that of comparison with earlier works. They proposed an alternative approach including the effect of
the average $\left\langle \vv{b}\right\rangle$. When starting the present work,
we did share the doubts of \citet{SBH:05} and were attracted by the alternative way of estimating
the fluctuating energy amount. But, having rejected them
in later stage, we feel obliged to explain our point of view.

Consider the example given by \citet{SBH:05}: a scalar stress $T= a_0 + a_1 \sin{t}$. There is no doubt that $\left\langle T^2 \right\rangle$ with $a_0 \neq 0 $
is different than for $a_0 = 0$. Moreover, with $a_0 \neq 0$ the plot of $U_\mathrm{e}(t)$ can be dominated by the $\sin{t}$, whereas dropping the
constant part we obtain only $\cos{2 t}$. However, a similar situation is met in a damped harmonic oscillator driven by $a_0 + a_1 \cos{t}$: the potential energy
of stationary solution has the same dependence on $a_0$, but the power loss is completely independent on $a_0$. The statement that only
a time variable part of stress may dissipate the energy can be found already in the paper of \citet{Prend:58}. There is no reason to doubt it.
Thus the only question is: does the the dissipation by periodic part of the stress depend on the constant part ? We cannot rule out such possibility
if relations between stress and strain are strongly nonlinear or the dissipation mechanism is complicated, but -- on the other hand -- it
is quite unlikely that in these circumstances a simple term $a_0 a_1 \sin{t}$ will properly describe the dependence, as \citet{SBH:05} suggest, mentioning
a friction due to grain boundary or crack surface sliding. So, the alternative estimate of the stored energy, as proposed by Sharma,
neither looks promising far from almost-linear, weakly damped elasticity model, nor it behaves properly within this area -- the
latter readily seen if  we consider the harmonic oscillator.
Moreover, it does not account for the fact, that a $k$-th harmonic in a general
harmonic load works $k$ times during the fundamental period -- the comment that applies
both the main and alternative solution of \citet{SBH:05}.  As a minor remark we can add the contradiction between qualifying
gravitation as an ignorable pre-stress and mean centrifugal force as a factor that contributes to the energy dissipation, present in \citet{SBH:05}.
The alternative recipe for $E_\mathrm{av}$ has a mathematical meaning, but in our opinion, it will define an alternative `$Q$ of a process'
which may be too far from the `intrinsic $Q$' in terms of numerical values and dependence or independence on parameters.

We believe that the main line of reasoning and the warnings issued by \citep{OCB:78} are sufficient to formulate the recipe for
$E_\mathrm{av}$ leading to a reasonable, material-based quality factor. The rule seems to be: stay close to the property that, for sufficiently high
$Q$ values,
\begin{equation} \label{Qfi}
Q^{-1} \approx \tan{\varphi},
\end{equation}
where $\varphi$ is the phase lag angle between the stress and the strain (so-called loss angle). This rule validates the separation of contributions from
subsequent harmonics of body forces $\vv{b}$, as well as the rejection of its constant part in the stress solution, i.e. the procedure of \citet{Efroimsky:2000}.
Note that, for a single harmonic, taking twice the mean value of its square we obtain its squared amplitude which explains why $2 E_\mathrm{av}$ in equation~(\ref{Qdef:1})
is often replaced by `peak energy' $E_\mathrm{p}$ \citep{OCB:78,Lambeck:80}.

Energy is dissipated by the work of body forces. The basic formula for the work rate $\dot{W}$ reads
\begin{equation}\label{wrk:1}
    \dot{W} = \int_V \dot{\epsilon} \,\mathrm{d}V = \int_V T_{ij} \dot{e}_{ij} \,\mathrm{d}V.
\end{equation}
In the conservative, elastic case, when stress and strain are defined by equations~(\ref{Tform}) and (\ref{Hooki}),
depending on time-periodic functions $B_{ij}$, the integral over the fundamental period of wobbling
renders no work, because each term of the sum $T_{ij} \dot{e}_{ij}$ is a purely periodic function of time.

\underline{Assumption 7:} Inelastic oscillations will be described by the following, heuristic approximation:
\begin{itemize}
\item Periodic terms of $B_{ij}$ in stress tensor $\mtr{T}$ are taken directly from the definition (\ref{Bdef:1}) and (\ref{lower}).
\item Periodic terms of $B_{ij}$ in strain tensor $\mtr{e}$, derived using equation~(\ref{Hooki}), are modified by adding a phase lag $\varphi$
to the argument of each harmonic, and dividing the amplitudes by $\cos{\varphi}$.
\item The phase lag is independent on coordinates $x$, $y$, $z$, and
related with the quality factor by equation~(\ref{Qfi}).
\item The quality factor is independent on the frequency of forcing terms.
\end{itemize}
Let the fundamental frequency of wobbling be $\Omega$, with an associated period
\begin{equation}\label{perio}
    P = \frac{2 \pi}{\Omega},
\end{equation}
and consider an exemplary product of periodic terms appearing in equation~(\ref{wrk:1}) under Assumption 7
\begin{eqnarray}
    p_{\mathrm{ex}} & = &  \left(c_p \cos{(p \Omega t)} + s_p \sin{(p \Omega t)}\right)
    \frac{\mathrm{d}~}{\mathrm{d}t}\left(\frac{c_q}{\cos{\varphi}} \cos{(q \Omega t-\varphi)} \right. \nonumber \\
    & & \left. + \frac{s_q}{\cos{\varphi}} \sin{(q \Omega t-\varphi)}\right).
\label{exe:1}
\end{eqnarray}
Straightforward computation leads to the conclusion, that the time integral over
the period $P$ vanishes for $p \neq q$, whereas $p=q \neq 0$ leads to
\begin{eqnarray}
   \int_0^P p_{\mathrm{ex}} \,  \mathrm{d}t  & = &     p \pi  \left(c^2_p + s_p^2\right) \tan{\varphi} = \nonumber \\
   & = &  p \frac{2 \pi}{Q} \left\langle  \left(c_p \cos{(p \Omega t)} + s_p \sin{(p \Omega t)}\right)^2 \right\rangle.
\label{exe:2}
\end{eqnarray}
This example establishes the link between Assumption 7 and the operational rule of computing
the energy loss due to the work per cycle $P$ as the sum
\begin{equation}\label{diss:fun}
   \Delta E = - \int_0^P \dot{W}\,\mathrm{d}t = - \frac{2 \pi}{Q} \sum_{p \geqslant 1} p \left\langle 2\, U_{p} \right\rangle,
\end{equation}
in agreement with equation~(\ref{Qdef:1}), where $U_p$ is the part of elastic energy $U_\mathrm{e}$ from equation~(\ref{elase})
involving only the $p$-th harmonic in each $B_{ij}$ term entering strain and stress.

An recent in-depth critical review of some issues concerning Assumption 7 can be found in \citep{Efro:2012}.

\section{Energy dissipation rate}
\label{Sec:4}
\subsection{Volume integration}

Energy dissipation rate can be obtained from equation~(\ref{diss:fun}) as
\begin{equation}\label{pow}
    \dot{E} = \frac{\Delta E}{P} = - \frac{2\,\Omega}{Q} \sum_{p \geqslant 1} p \left\langle  U_{p} \right\rangle.
\end{equation}
Having assumed the independence of $\varphi$ and $Q$ on coordinates, we can perform the volume integration
required for $U_p$ before the time average. This allows a considerable economy of expressions; substituting
the general form of $\mtr{T}$ from equation~(\ref{Tform}) into equation~(\ref{edens}), making use of
the expressions for $\mtr{A}$, $\mtr{A}^{ij}$ from Appendix~\ref{ap:A}, plugging in the definition of $\mtr{B}$
in (\ref{Bdef:1},\ref{lower}), and integrating over the ellipsoid volume according to equation~(\ref{vol:int}),
we find
\begin{eqnarray}
  \left\langle  U_{p} \right\rangle &=& \frac{a^4 \rho \, m}{\mu} \left(
  \alpha_{11} \left\langle \left[\omega_1^2\right]^2_p \right\rangle
  + \alpha_{22} \left\langle \left[\omega_2^2\right]^2_p \right\rangle
   + \alpha_{33} \left\langle \left[\omega_3^2\right]^2_p \right\rangle  \right.\nonumber \\
   & &+\alpha_{12} \left\langle \left[\omega_1^2\right]_p \left[\omega_2^2\right]_p\right\rangle
+\alpha_{13} \left\langle \left[\omega_1^2\right]_p \left[\omega_3^2\right]_p\right\rangle  \nonumber \\
 & & +\alpha_{23} \left\langle \left[\omega_2^2\right]_p \left[\omega_3^2\right]_p\right\rangle
 + \beta_{12} \left\langle \left[\omega_1 \omega_2\right]^2_p \right\rangle \nonumber \\
 & & \left. + \beta_{13} \left\langle \left[\omega_1 \omega_3\right]^2_p \right\rangle
+ \beta_{23} \left\langle \left[\omega_2 \omega_3\right]^2_p \right\rangle \right),
\label{Upav}
\end{eqnarray}
where $m$ is the ellipsoid mass and $\alpha_{ij}$, $\beta_{ij}$ are dimensionless rational functions
of $h_1$, $h_2$ and $\nu$.
For any function $F$ represented as Fourier series, the symbol $[F]_p$ designates its $p$-th harmonic, i.e. a
trigonometric monomial with argument $p \Omega t$.

Regretfully, the full form of $\alpha_{ij}$ and $\beta_{ij}$ is too long to be explicitly quoted
(although short enough to be efficiently programmed), but the expressions are available
from the authors in an electronic form.

\subsection{Fourier harmonics of angular velicity}

Equation (\ref{Upav}) requires the recall of basic facts about the free rotation of rigid body  \citep{Whit:52}
for the specific case of a homogenous ellipsoid.  As usually, we have to distinguish the Short Axis Mode (SAM) of rotation, when
angular velocity vector $\vv{\omega}$ circulates around $\vv{e}_3$, and the Long Axis Mode (LAM), when $\vv{\omega}$
circulates around $\vv{e}_1$. Quantities referring to the SAM will have subscript $s=3$, and those for the LAM -- subscript $s=1$
(in this paper, we use $s$ exclusively for labeling the rotation mode).

Using two invariants of the free top, i.e kinetic energy $K_0$ and angular momentum $\vv{H}$ (equations
(\ref{K0}) and (\ref{angmom})),
with diagonal matrix of inertia $\mtr{I}$ given by equation~(\ref{iner}),
we define an auxiliary quantity $\mathcal{A}$ \citep{DepEl:93, BRV:2011}
\begin{equation}\label{Adef}
    \mathcal{A} = \frac{2 K_0 }{H^2},
\end{equation}
such that $a_3 \leqslant \mathcal{A} \leqslant a_1$, where
\begin{equation}
    a_i = I^{-1}_i.
\end{equation}

A nominal angular velocity of rotation $H a_3$ is often adopted as a scaling factor in the energy dissipation models.
Its value would be equal to $\omega$ only in the principal axis rotation around $\vv{e}_3$. We prefer to use a
quantity appropriate for the LAM as well, so let us define the nominal angular rates as
\begin{equation}\label{omnom}
    \tilde{\omega}_j = H a_j,
\end{equation}
such that each  $ \tilde{\omega}_j = \omega$  in the principal axis rotation around $\vv{e}_j$.

Let us define
\begin{equation}
 n_3 =    \sqrt{(a_1-\mathcal{A})(a_2-a_3)}, \qquad n_1 =    \sqrt{(a_1-a_2)(\mathcal{A}-a_3)}.
\end{equation}
Then, the components of angular velocity $\omega_i$ in the body fixed frame are expressible in terms of
the Jacobi elliptic functions with argument $\tau_s = H n_s\,t$ and modulus $k_s$, where
\begin{equation}\label{k13}
    k_3 =  \frac{n_1}{n_3} = \frac{1}{k_1}.
\end{equation}
Thus, in any rotation mode,
\begin{eqnarray}
  \omega_1  &=& \tilde{\omega}_1 \sqrt{\frac{\mathcal{A}-a_3}{a_1-a_3}}\, F^{(1)}_s, \nonumber  \\
  \omega_2  &=& \tilde{\omega}_2 \sqrt{\frac{\mathcal{A}-a_3}{a_2-a_3}}\, F^{(2)}_s, \\
  \omega_3  &=& \tilde{\omega}_3 \sqrt{\frac{a_1-\mathcal{A}}{a_1-a_3}}\, F^{(3)}_s,  \nonumber
\end{eqnarray}
with specific functions
\begin{equation}
F^{(1)}_3 = \pm \mathrm{cn}(\tau_3,k_3), \quad F^{(2)}_3 = \mathrm{sn}(\tau_3,k_3),
 \quad F^{(3)}_3 = \pm \mathrm{dn}(\tau_3,k_3),
\end{equation}
and
\begin{equation}
F^{(1)}_1 = \pm \mathrm{dn}(\tau_1,k_1), \quad F^{(2)}_1 = \pm  k_1 \, \mathrm{sn}(\tau_1,k_1),
 \quad F^{(3)}_1 = \mathrm{cn}(\tau_1,k_1),
\end{equation}
for the SAM and LAM, respectively.

The fundamental frequency of wobbling, appearing in equation~(\ref{perio}) as $\Omega$, is
\begin{equation}\label{om:s}
    \Omega_s = \frac{\pi H  n_s}{2 \mathrm{K}_s},
\end{equation}
where $\mathrm{K}_s $ stands for the complete elliptic integral of the first kind
\begin{equation}\label{Kdef}
    \mathrm{K}_s = \mathrm{K}(k_s) = \int_0^\frac{\pi}{2} \frac{\mathrm{d}x}{ \sqrt{1- k_s^2 \sin^2{x}}}.
\end{equation}
Similarly, we will use $\mathrm{E}_s = \mathrm{E}(k_s)$ for the complete elliptic integral of the second kind.

In the present work, we are interested only in the squares and products of angular velocity components.
Given the Jacobi elliptic functions with argument $\tau_s$ and modulus $k_s$, we can expand their squares and
products in Fourier series of angle
\begin{equation}\label{psitau}
    \psi_s = \frac{\pi}{2 \mathrm{K}_s}\,\tau_s = \Omega_s \, t.
\end{equation}
From the expressions available in \citet{BF:1954}, we can easily derive series
\begin{eqnarray}
  \mathrm{sn}^2(\tau_s,k_s) &=& X_0 - \sum_{p = 1}^\infty X_{2p} \cos{2p\psi_s}, \\
  \mathrm{cn}^2(\tau_s,k_s) &=& 1-X_0 + \sum_{p = 1}^\infty X_{2p} \cos{2p\psi_s}, \\
  \mathrm{dn}^2(\tau_s,k_s) &=& 1- k_s^2 X_0 + k_s^2 \sum_{p = 1}^\infty X_{2p} \cos{2p\psi_s},\\
  \mathrm{sn}(\tau_s,k_s) \, \mathrm{cn}(\tau_s,k_s) &=&  \sum_{p = 1}^\infty Y_{2p} \sin{2p\psi_s},\\
  \mathrm{cn}(\tau_s,k_s) \, \mathrm{dn}(\tau_s,k_s) &=& k_s \sum_{p = 1}^\infty X_{2p-1} \cos{(2p-1)\psi_s},\\
  \mathrm{sn}(\tau_s,k_s) \, \mathrm{dn}(\tau_s,k_s) &=& k_s \sum_{p = 1}^\infty Y_{2p-1} \sin{(2p-1)\psi_s},
\end{eqnarray}
with coefficients
\begin{eqnarray}
  X_0 &=& \frac{1}{k^2_s} \left( 1 - \frac{\mathrm{E}_s}{\mathrm{K}_s} \right), \\
  X_j &=& \left(\frac{\pi}{k_s\, \mathrm{K}_s}\right)^2 \,\frac{j \,\mathrm{q}_s^{j/2}}{1- \mathrm{q}_s^j},\\
  Y_j &=& \frac{1- \mathrm{q}_s^j}{1 + \mathrm{q}_s^j}\, X_j,
\end{eqnarray}
involving the Jacobi's nome
\begin{equation}\label{qdef}
    \mathrm{q}_s = \exp{\left( - \pi \mathrm{K}'_s  / \mathrm{K}_s  \right)},
\end{equation}
where $\mathrm{K}'_s = \mathrm{K}(k'_s)$ is the elliptic integral with complementary modulus
\begin{equation}
k'_s = \sqrt{1-k_s^2}.
\end{equation}
Quickly convergent series for the nome
\begin{equation}\label{qapr}
   \mathrm{q}_s = \zeta_s  \left( 1 +2 \zeta_s^4 + 15\,\zeta_s^8 + 150 \, \zeta_s^{12} +1707 \, \zeta_s^{16}+ \ldots \right),
\end{equation}
with
\begin{equation}\label{zeta}
    \zeta_s = \frac{1- \sqrt{k'_s}}{2\,\left( 1+\sqrt{k'_s}\right)},
\end{equation}
can be used \citep{Innes:1902,BF:1954}. Even close to the separatrix $(\mathcal{A} \approx a_2)$, the first five terms
provide a relative error of $10^{-6}$ for $k_s = 0.999$.
However, for the values of $k_s$ close to 1, it is better to use
\begin{equation}\label{zetapr}
    \zeta'_s = \frac{1- \sqrt{k_s}}{2\,\left( 1+\sqrt{k_s}\right)},
\end{equation}
in the right-hand side of (\ref{qapr}) instead of $\zeta$, and obtaining the complementary nome $\mathrm{q}'_s$, which
may serve to compute $\mathrm{q}_s$ through the relation
\begin{equation}\label{qqp}
    \ln \mathrm{q}_s \, \ln \mathrm{q}'_s = \pi^2.
\end{equation}
An additional benefit of the nome is also a quickly convergent series for the elliptic integral
\begin{equation}\label{Kser}
    \mathrm{K}_s = \frac{\pi}{2} \left( 1+ 2 \sum_{p=1}^\infty \mathrm{q}_s^{p^2}\right)^2.
\end{equation}

The mean values required in equation~(\ref{Upav}) follow directly from the presented Fourier series.
For example, in the SAM
\begin{eqnarray}
\left\langle \left[\omega_1^2\right]^2_{2p} \right\rangle & =  &
 \tilde{\omega}_1^4 \,\left(\frac{\mathcal{A}-a_3}{a_1-a_3}\right)^2 \,\left\langle \left(X_{2p} \cos{2p\psi_3} \right)^2  \right\rangle =
\nonumber \\
& = & \frac{\tilde{\omega}_1^4 }{2}\,\left(\frac{\mathcal{A}-a_3}{a_1-a_3}\right)^2 \,X_{2p}^2,  \label{examp:1} \\
  \left\langle \left[\omega_1^2\right]^2_{2p-1} \right\rangle & = & 0,
\end{eqnarray}
and in the LAM
\begin{equation}\label{examp:2}
\left\langle \left[\omega_1^2\right]^2_{2p} \right\rangle = \frac{\tilde{\omega}_1^4}{2}
\, \left(\frac{\mathcal{A}-a_3}{a_1-a_3}\right)^2 \, k_1^4 X_{2p}^2,
\qquad \left\langle \left[\omega_1^2\right]^2_{2p-1} \right\rangle = 0,
\end{equation}
with $X_{2p}$ depending on $\mathrm{q}_3$ and $\mathrm{q}_1$, respectively.
Note, that some of the mean values can be negative, like
\begin{equation}
\left\langle \left[\omega_1^2\right]_{2p} \left[\omega_2^2\right]_{2p} \right\rangle
= - \frac{\tilde{\omega}_1^2 \tilde{\omega}_2^2 }{2}
\,  \frac{\left(\mathcal{A}-a_3\right)^2}{\left(a_1-a_3\right)\left(a_2-a_3\right)} \,  X_{2p}^2,
\end{equation}
in the short axis mode. However, their associated $\alpha_{ij}$ are also negative, so
there is no subtraction in equation~(\ref{Upav}).

\subsection{Final expressions}

Performing the necessary substitutions in equation~(\ref{pow}), we find
an expression for the energy loss rate
\begin{equation}
  \dot{E}_s = - \frac{a^4 \rho \, m \tilde{\omega}_s^5}{\mu\, Q} \, \Psi_s(k_s,h_1,h_2,\nu) \label{dEs}
\end{equation}
with dimensionless
\begin{eqnarray}
  \Psi_3  & = & Z_3^5 \left( P_1(k_3) M_{13} +P_2(k_3) M_{23}  +P_3(k_3)  M_0 + P_4(k_3) M_{12} \right), \nonumber \\
 & &   \label{dE3} \\
  \Psi_1  & = & Z_1^5 \left( P_1(k_1) M_{13} +P_2(k_1) M_{12}  + P_3(k_1)M_0  + P_4(k_1) M_{23}  \right), \nonumber \\
  & &  \label{dE1}
\end{eqnarray}
(note the swapped $M_{23}$ and $M_{12}$), where
\begin{equation}
   Z_s   =   \frac{\Omega_s}{\tilde{\omega}_s} = \frac{\pi  n_s}{2 a_s \mathrm{K}_s}.
\end{equation}
First, we recall that the leading factor
depends on the semi-major axis of ellipsoid $a$, its mass $m$, density $\rho$,  the fifth power of the nominal
rotation rate $\tilde{\omega}_s$ (resulting from the division of angular momentum $H$ by the related
moment of inertia), on Lam\'{e} shear modulus $\mu$, and
quality factor $Q$. Functions $P_i(k_s)$ depend on the ratio of kinetic energy and angular momentum through
an elliptic modulus $k_s$ that enters the Jacobi's nome $\mathrm{q}_s$, and have the form of infinite sums
\begin{eqnarray}
  P_1(k_s) &=& \sum_{p=1}^\infty \frac{(2p-1)^3 \mathrm{q}_s^{2p-1}}{\left(1-\mathrm{q}_s^{2p-1}\right)^2}, \label{P:1}\\
  P_2(k_s) &=& \sum_{p=1}^\infty \frac{(2p-1)^3 \mathrm{q}_s^{2p-1}}{\left(1+\mathrm{q}_s^{2p-1}\right)^2}, \\
  P_3(k_s) &=& \sum_{p=1}^\infty \frac{(2p)^3 \mathrm{q}_s^{2p}}{\left(1-\mathrm{q}_s^{2p}\right)^2}, \\
  P_4(k_s) &=& \sum_{p=1}^\infty \frac{(2p)^3 \mathrm{q}_s^{2p}}{\left(1+\mathrm{q}_s^{2p}\right)^2}, \label{P:4}
\end{eqnarray}
although in practice only a few leading terms should be sufficient.
Finally, $M_{ij}$ and $M_0$ are dimensionless, positive coefficients depending only on the shape,
(through $h_1$, $h_2$) and on the Poisson's ratio $\nu$. In terms of the coefficients from equation~(\ref{Upav}), they are
\begin{eqnarray}
  M_{ij} & = & \frac{16 \,a_i^2 a_j^2 \beta_{ij}}{d_{12} d_{13}  d_{23} d_{ij}}, \\
  M_0 & = &  \frac{16 }{d_{12} d_{13}  d_{23}} \left(
  \frac{a_1^4 d_{23} \alpha_{11}}{d_{12} d_{13}}+\frac{a_2^4 d_{13} \alpha_{22}}{d_{12} d_{23}}
  +\frac{a_3^4 d_{12} \alpha_{33}}{d_{13} d_{23}}
  \right. \nonumber \\
  & & \left. - \frac{a^2_1 a^2_2 \alpha_{12}}{d_{12}} + \frac{a^2_1 a^2_3 \alpha_{13}}{d_{13}}
  - \frac{a^2_2 a^2_3 \alpha_{23}}{d_{23}} \right).
\end{eqnarray}
where $d_{ij}=(a_i-a_j)$.
Appendix~\ref{ap:B} contains the full expressions of $M_{ij}$ and $M_0$. For the reasons explained in the next section,
we give them with the fixed  Poisson's ratio $\nu = 0.25$.

\subsection{Poisson's ratio}

All recent models of spin axis relaxation assume the Poisson's ratio $\nu = 0.25$, i.e.
equal Lam\'{e} constants $\lambda = \mu$. Authors justify it by the fact, that this is approximately a typical value
for most of cold solids \citep{Efroimsky:2000}. Earlier, \citet{Prend:58} considered an incompressible object
with $\nu = 0.5$. Only \citet{MoMo:03} maintain the explicit dependence on $\nu$ in their final formulae for a spheroid.

The present model also maintains $\nu$ in the final expressions, so we are in a favorable situation to estimate the sensitivity of $\dot{E}_s$
on its value. Interestingly, in contrast to the results of \citet{MoMo:03},
the dependence of $M_{ij}$ and $M_0$ on the Poisson's occurs to be very weak.
As a function of $0 \leqslant \nu \leqslant 0.5$, the values of $M$ coefficients vary on the level of at most $10^{-2}$
(relatively), whereas the solution of \citet{MoMo:03} exhibits the dependence on the level of $10^{-1}$.
This property came unexpected, because $\alpha_{ij}$ still contained a factor $(1-\nu^2)^{-1}$,
that later vanished in $M_0$. In these circumstances, we fix the value of $\nu = 0.25$ as a physically realistic one,
which considerably simplifies expressions, but the results will fairly well apply to an incompressible case with $\nu = 0.5$.

\section{Wobble damping time}

\label{Sec:5}

Let us define a `wobbling angle' $\theta_s$ as the maximum angle between
the angular momentum vector $\vv{H}$ and a relevant axis
($Oz$ in SAM, or $Ox$ in LAM) attained during the wobbling cycle of a rigid body, namely
\begin{equation}\label{thdef}
    \theta_s = \max \left(\arccos \left| \frac{\vv{H}\cdot \vv{e}_s}{H} \right|\right).
\end{equation}
In the principal axis rotation $\theta_s =0$, regardless of prograde or retrograde case.
The other limit, $\theta_s = 90^\circ$, refers to unstable rotation around an intermediate axis of inertia
$Oy$, or to the nonperiodic rotation on the separatrix, which falls beyond our model.

There is a direct relation between $\theta_s$ and the variable $\mathcal{A}$
\begin{equation}\label{Ak}
    \mathcal{A}_s = a_2 - \left(a_2-a_s\right) \cos^2{\theta_s}.
\end{equation}
We have introduced the  subscript $s$ to $\mathcal{A}$
in order to facilitate the distinction of modes, although the primary definition
(\ref{Adef}) is universal.
The modulus $k_s$ is also related to the wobbling angle through
\begin{eqnarray}
  k_s &=& \frac{\sin\theta_s}{\sqrt{1+\kappa_s \cos^2\theta_s}}, \label{ksth} \\
  \kappa_3 &=& \frac{1}{\kappa_1} = \frac{a_2-a_3}{a_1-a_2} = h_1^4 \,\frac{1-h_2^4}{ 1-h_1^4 }.
\end{eqnarray}

Since the dissipation of energy does not affect angular momentum, and the energy is drained from the kinetic
$K_0$, the differentiation of equation~(\ref{Adef}) leads to
\begin{equation}
    \dot{\mathcal{A}}_s = \frac{2\,\dot{E}_s}{H^2} = - \frac{2 a^4 \rho \, m \tilde{\omega}_s^3}{\mu\, Q} \,a_s^2 \Psi_s ,
\end{equation}
where $\dot{E}_s$ has been taken from  equation~(\ref{dEs}).

On the other hand,
\begin{equation}
    \dot{\mathcal{A}}_s =  2\,\left(a_2-a_s\right) \sin{\theta_s} \cos{\theta_s} \dot{\theta}_s,
\end{equation}
and we can equate the two relations, obtaining the differential equation
\begin{equation}
    \frac{\mathrm{d}\theta_s}{\mathrm{d}t} = \frac{ \dot{E}_s}{H^2 \left(a_2-a_s\right) \sin{\theta_s} \cos{\theta_s}}.
\end{equation}
The resulting quadrature gives the time $T_s$ required to change wobbling angle from the initial $\theta^0_s$ to the final
$\theta'_s$
\begin{equation}\label{Ts}
    T_s = \frac{ \left(a_s-a_2\right) \mu Q}{a^4 \rho m \tilde{\omega}_s ^3 a_s^2 } \int_{\theta^0_s}^{\theta'_s}
    \frac{\sin{\theta_s} \cos{\theta_s}}{\Psi_s}\,\mathrm{d}\theta_s,
\end{equation}
where $\Psi_s$ should be expressed in terms of the wobbling angle using equation~(\ref{ksth}).

In particular,
\begin{equation}\label{T3}
    T_3 =  - \frac{  \mu Q}{a^2 \rho \tilde{\omega}_3 ^3} \,\left[
     \frac{h_1^2 \left(1+h_1^2\right)\left(1-h_2^2\right)}{5\,\left( 1+h^2_1 h^2_2 \right)} \right]
     \int_{\theta^0_3}^{\theta'_3}
    \frac{\sin{\theta_3} \cos{\theta_3}}{ \Psi_3}\,\mathrm{d}\theta_3,
\end{equation}
implying $\theta'_3 \leqslant \theta^0_3$, and
\begin{equation}\label{T1}
    T_1 =   \frac{  \mu Q}{a^2 \rho \tilde{\omega}_1 ^3}\,\left[
     \frac{h_1^2 \left(1-h_1^2\right)\left(1+h_2^2\right)}{5 \left(1+h^2_1 h^2_2\right)} \right]
     \int_{\theta^0_1}^{\theta'_1}
    \frac{\sin{\theta_1} \cos{\theta_1}}{ \Psi_1}\,\mathrm{d}\theta_1,
\end{equation}
with $\theta^0_1 \leqslant \theta'_1$.
For the reference with earlier works, we will use a shape parameter $D_s$ \citep{SBH:05} defined as
\begin{equation}\label{Ddef}
    D_s(h_1,h_2) = T_s \frac{a^2 \rho \tilde{\omega}_s ^3}{\mu Q},
\end{equation}
for prescribed integration limits.

Of course, the notion of wobble damping time is properly related to $T_3$ or, if the evolution starts in the LAM and continue through the SAM,
to the sum $T_1+T_3$. In the long axis mode, energy dissipation excites wobbling, driving the angular momentum vector towards the separatrix.
A term `excitation time' seems more appropriate for $T_1$.

\begin{figure}
 \begin{center}
  \includegraphics[width=8.5cm]{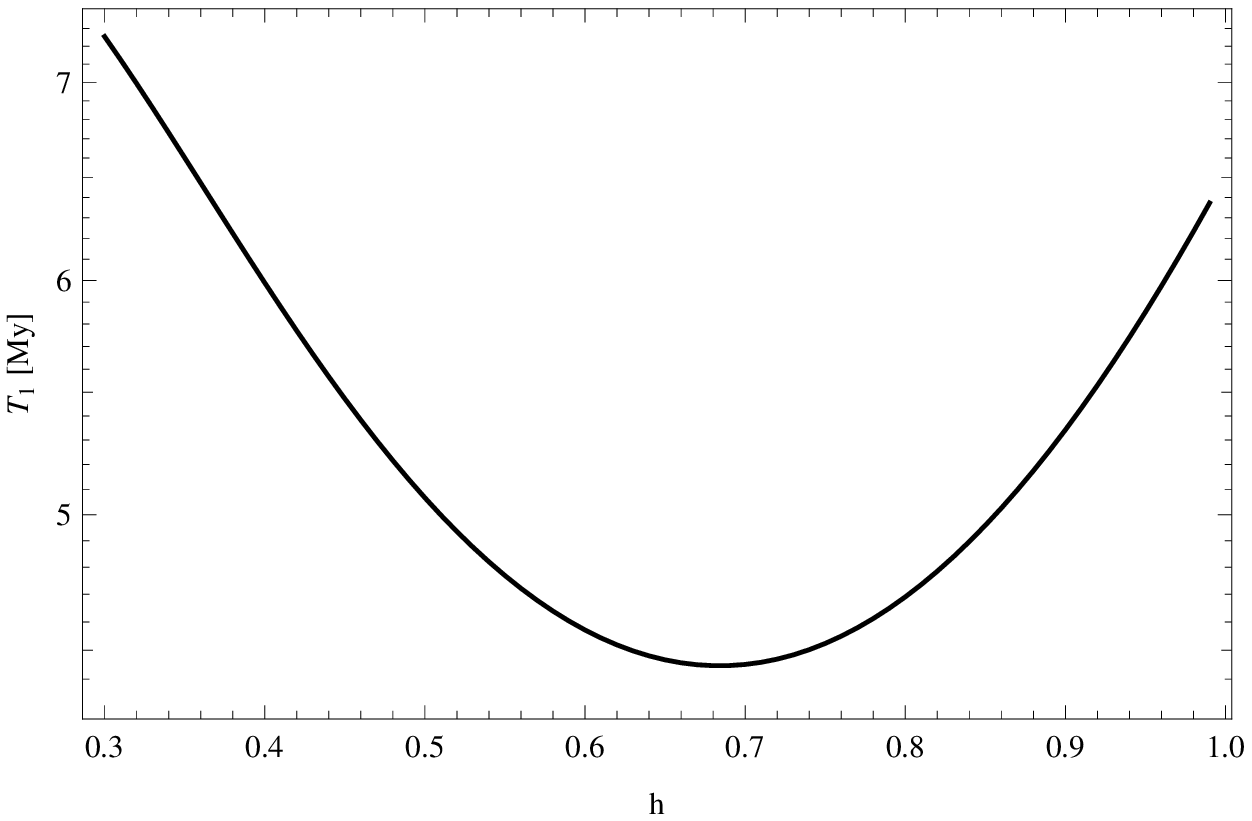}\\
  \includegraphics[width=8.5cm]{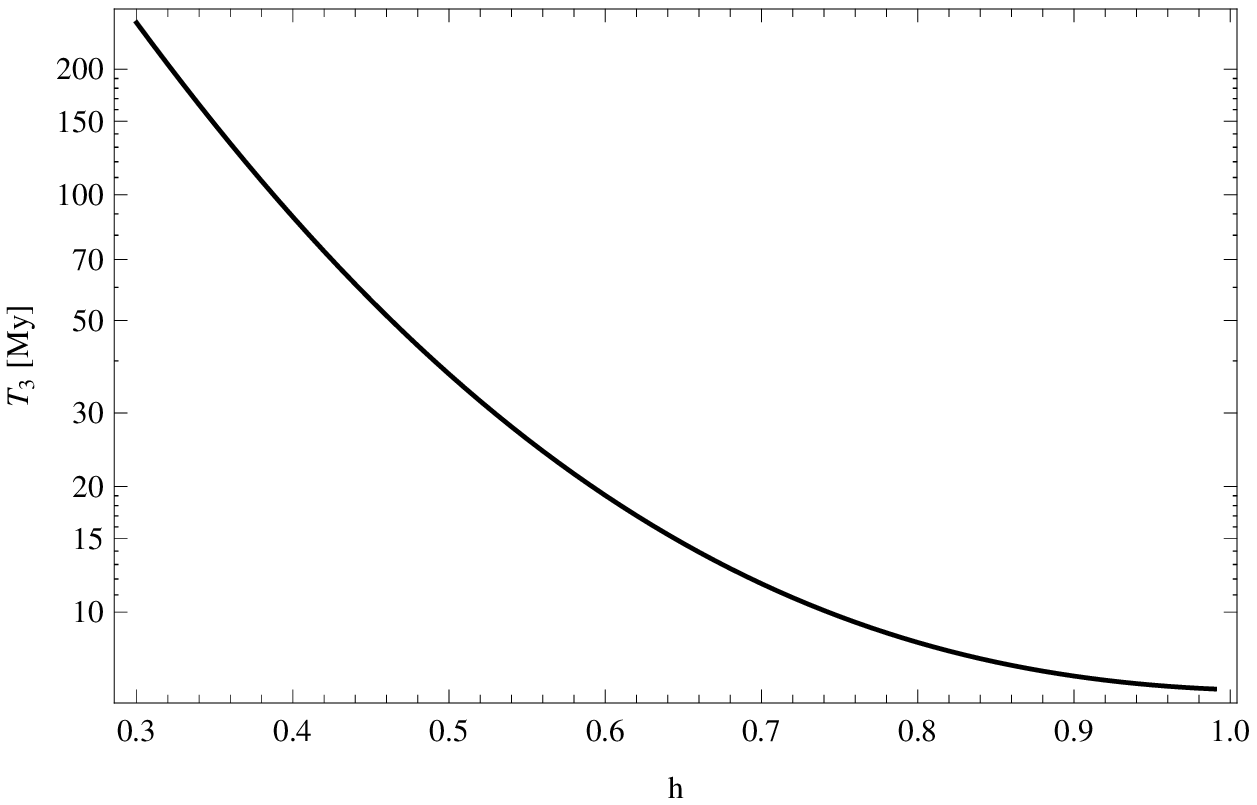}
  \end{center}
  \caption{Logarithmic plots of wobble damping/excitation times for ellipsoids with $h_1=h_2=h$.
  Top: time spent in the LAM with $\theta_1$ increasing from $5^\circ$ to $85^\circ$ \textbf{(corrected)};
  bottom: time spent in the SAM with $\theta_3$ decreasing from $85^\circ$ to $5^\circ$.
  Physical parameters -- see the text. }
  \label{fig:1}
\end{figure}
As an illustrative example, we first plot damping/excitation times for a family of ellipsoids with $h_1=h_2=h$, assuming sample physical data
$a = 1~\mathrm{km}$, $\rho = 2000~\mathrm{kg\,m^{-3}}$, $\mu = 10^9~\mathrm{Pa}$, and $Q=100$, chosen more for an ease of scaling
than for the reference to some specific case. We have assumed $\tilde{\omega}_3 = 2 \pi/10^\mathrm{h}$ for $T_3$
and computed an equivalent
\begin{equation}\label{omt13}
    \tilde{\omega}_1 = \frac{1+h_1^2}{h_1^2 \, \left(1+h_2^2 \right)}  \tilde{\omega}_3,
\end{equation}
to be used in $T_1$.

Interested in a possibly complete history, we start the evolution at the LAM, with $\theta_1^0=5^\circ$
and integrate equation~(\ref{T1}) up to $\theta_1' = 85^\circ$. The results are displayed in Fig.~\ref{fig:1} (top).
The dependence of $T_1$ on $h$ is not monotonous: the shortest excitation time, $4.4~\mathrm{My}$, occurs
at $h = 0.68$ \textbf{(corrected)}. Increasing asphericity, we reach $T_1 \approx 7.25~\mathrm{My}$  at $h=0.3$ \textbf{(corrected)}. The other extreme would be $h=1$,
but we stop at $h=0.99$, because the results would be meaningless for a sphere. After crossing the separatrix,
the angular momentum vector is driven towards $\vv{e}_3$ and we computed the damping times $T_3$ taken
to evolve from $\theta_3^0 = 85^\circ$ to $5^\circ$. This time (Fig.~\ref{fig:1}, bottom),
damping times are much longer than in the LAM for the same shape. $T_3$ is as high as $258~\mathrm{My}$ at $h=0.3$ and
systematically decreases to $6.5~\mathrm{My}$ at $h=0.99$. Thus,  the total damping time consists mostly in $T_3$ and --
fixing the semi-axis $a$ -- we find that triaxiality inhibits the total damping process, save for a quick
passage through the LAM, where we find a minimum at $h_1=h_2=0.75$.

\begin{figure}
 \begin{center}
  \includegraphics[width=8.5cm]{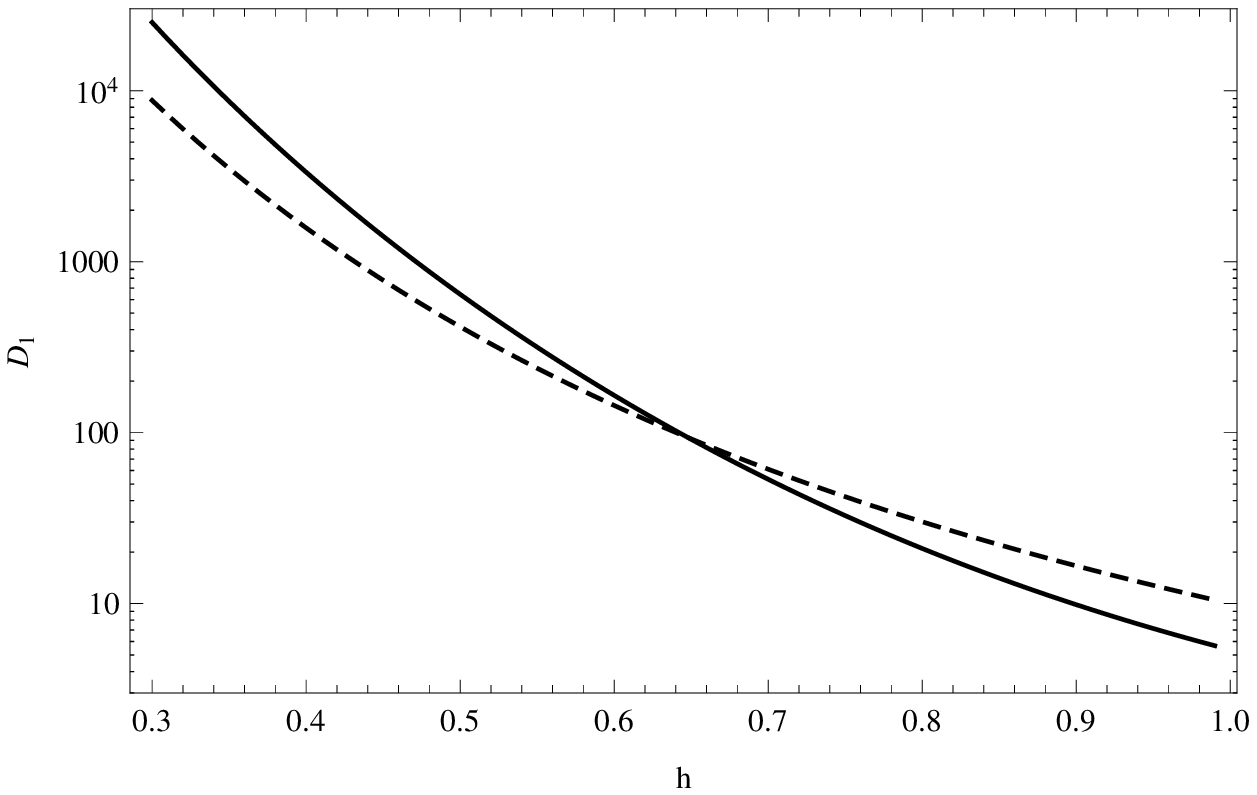}\\
  \includegraphics[width=8.5cm]{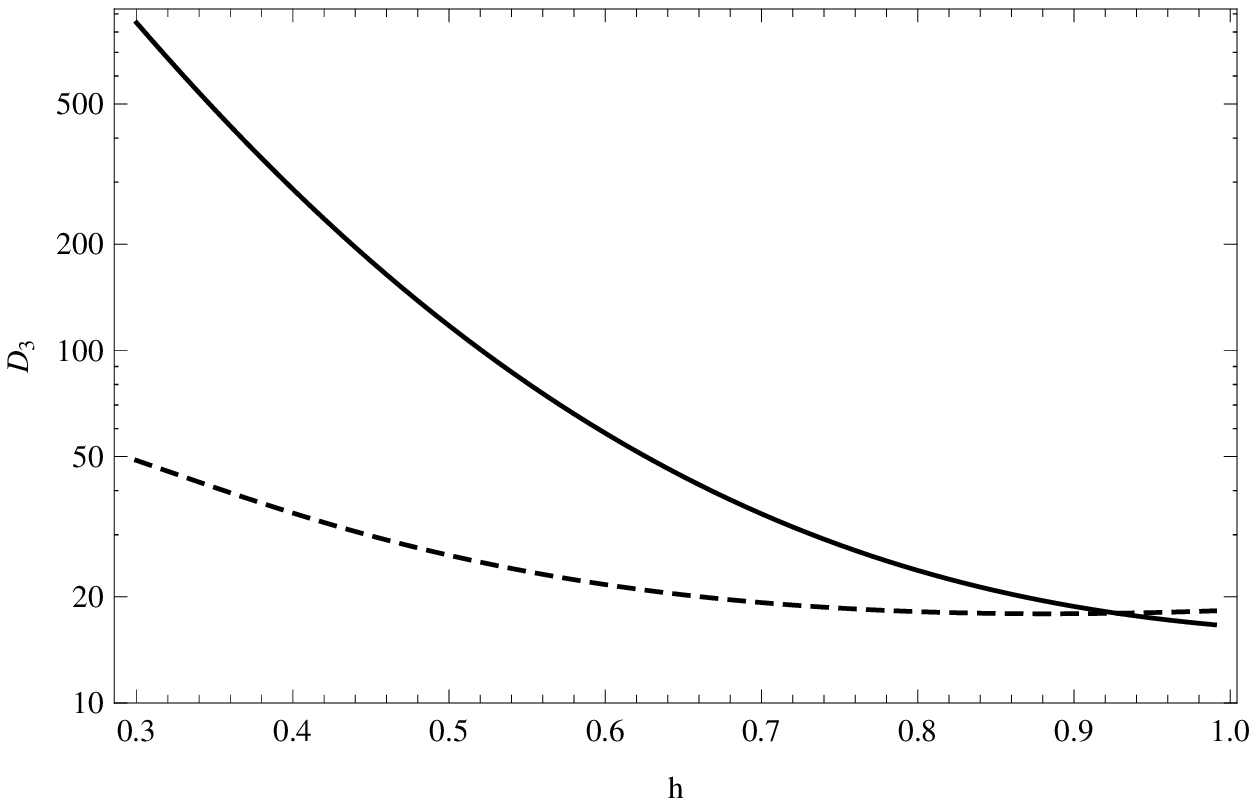}
  \end{center}
  \caption{Logarithmic plots of shape functions $D_s$. Solid line: ellipsoids with $h_1=h_2=h$;
  dashed line: spheroids with $h_1=h$ (LAM -- top), or with $h_2=h$ (SAM -- bottom).}
  \label{fig:2}
\end{figure}

Integration limits in the above example have been wider than usually. For the reference with earlier works
we have also computed shape factors $D_s$ for the limits $45^\circ \leqslant \theta_1 \leqslant 85^\circ$, and
$45^\circ \geqslant \theta_3 \geqslant 5^\circ$, like \citet{SBH:05}. Figure~\ref{fig:2} confronts $D_s$ for
a family of $h_1=h_2=h$ ellipsoids with spheroids having an appropriate ratio $h_1$ (SAM) or $h_2$ (LAM) equal to 1,
and the other one set as $h$.
We note a systematic increase of $D_s$ with increasing $h$ for ellipsoids,similar to $T_3$, but unlike $T_1$ from
Fig.~\ref{fig:1}. For oblate spheroids there is shallow minimum of $D_3$ close to $h \approx 0.9$.
These discrepancies are not essential and result from an arbitrary choice of integration limits that cut off
some more or less prominent (depending on the shape) parts of an integrand.

\begin{figure}
 \begin{center}
  \includegraphics[width=8.5cm]{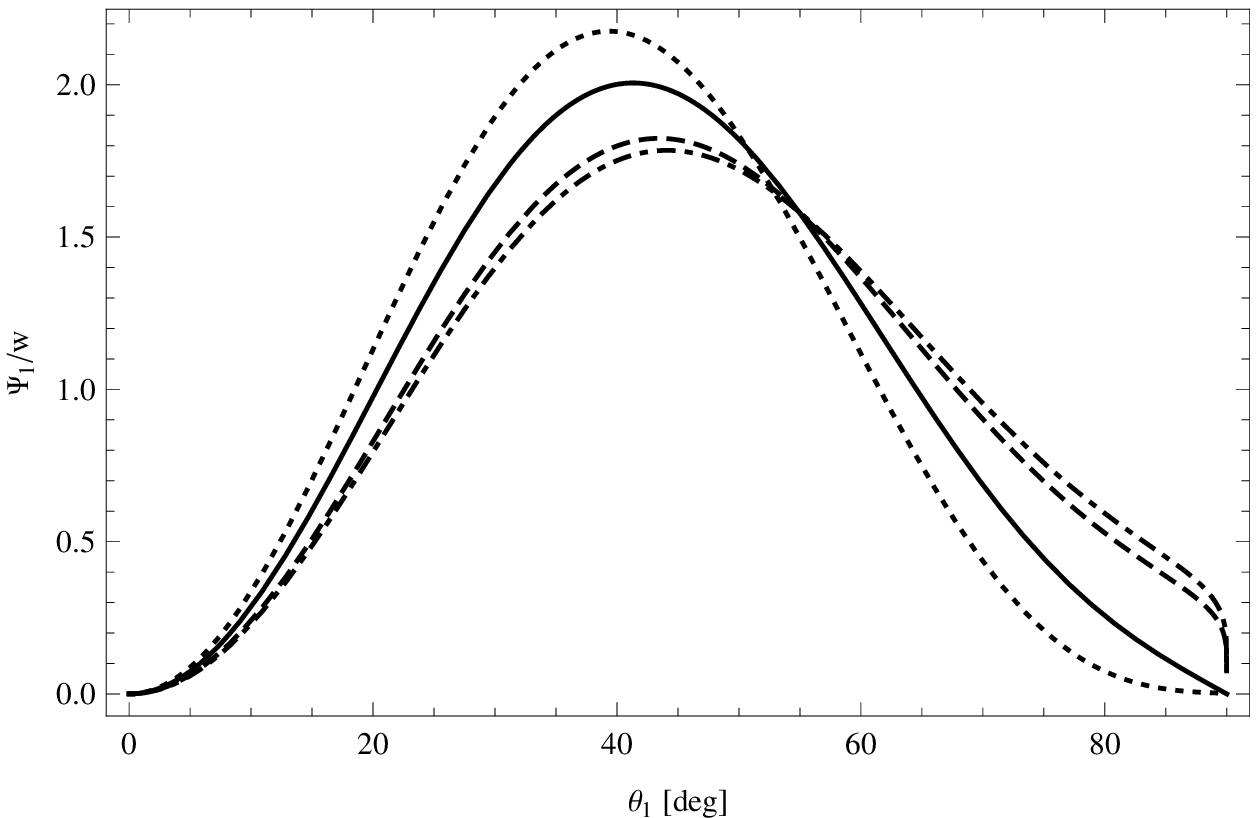}
  \end{center}
  \caption{Normalized LAM energy dissipation rate functions $\Psi_1(\theta_1)/w$.
  Solid line: $(h_1,h_2,1/w) = (0.7,1, 3000)$; dashed: $(0.7,0.7,4000)$; dotted: $(0.3,0.7, 9.6 \times 10^5)$;
  dot-dashed: $(0.7,0.3,5200)$. \textbf{(corrected)}}
  \label{fig:3}
\end{figure}

\begin{figure}
 \begin{center}
  \includegraphics[width=8.5cm]{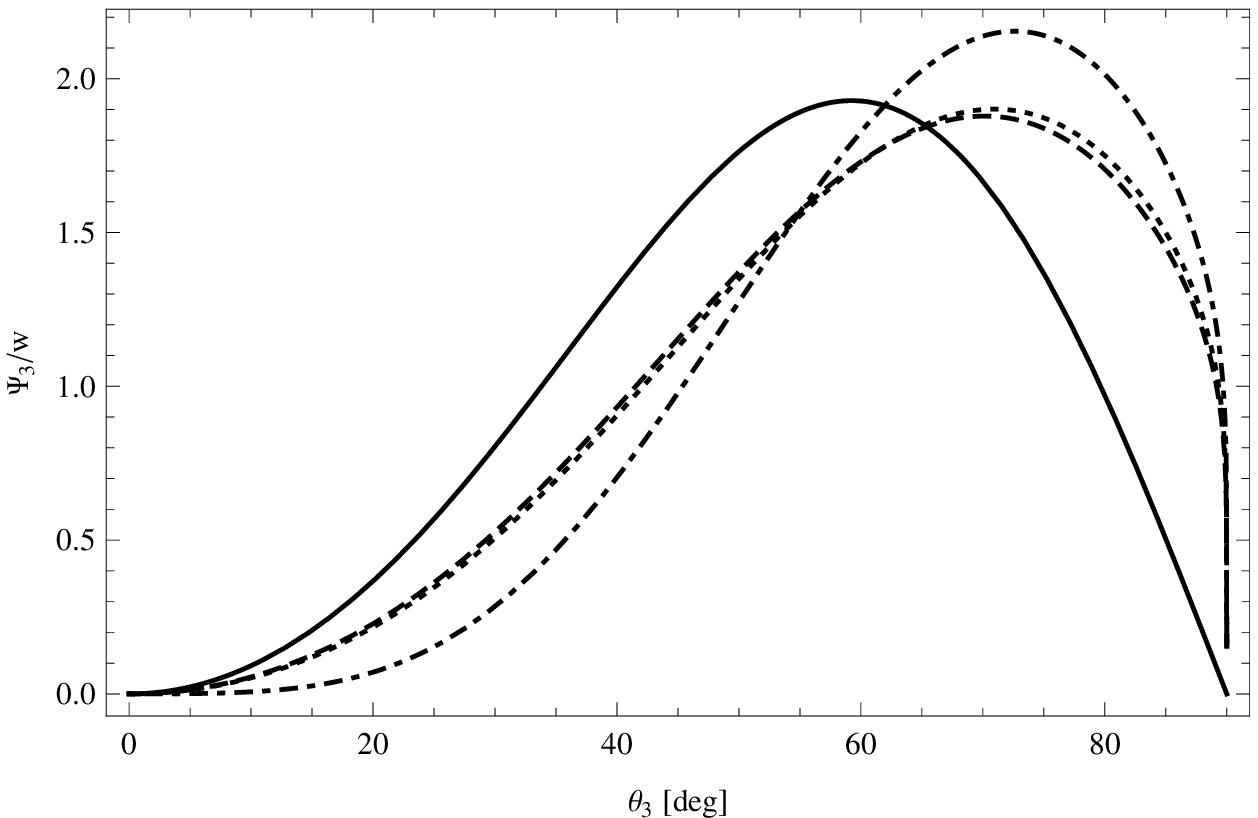}
  \end{center}
  \caption{Normalized SAM energy dissipation rate functions $\Psi_s(\theta_s)/w$.
  Solid line: $(h_1,h_2,1/w) = (1,0.7, 210)$; dashed: $(0.7,0.7,530)$; dotted: $(0.3,0.7, 1.6 \times 10^4)$;
  dot-dashed: $(0.7,0.3,140)$. \textbf{(corrected)}}
  \label{fig:4}
\end{figure}

Integrated damping/excitation time is important for qualitative considerations, but if the joint action of
inelastic dissipation and other torques is to be studied, the shape of $\Psi_s(\theta_s)$ becomes more interesting.
Figures~\ref{fig:3} and \ref{fig:4} demonstrate, how the ellipsoid's shapes
affect $\Psi_1(\theta_1)$ and $\Psi_3(\theta_3)$.
Each curve is normalized, i.e. $\Psi_s$ values are divided by $w$  -- the mean value of
$\Psi_s$ with respect to $\theta_s$ on the interval  $[0,\pi/2]$.
The situation is a bit different in the LAM (Fig.~\ref{fig:3}) and SAM (Fig.~\ref{fig:4}).
Departure from a spheroid (solid line) weakens the dissipation close to the principal axis
and amplifies it  for higher $\theta_3$ in the SAM. In the LAM, this is not the rule, as seen
for $h_1=0.3$ and $h_2=0.7$.

\section{Reduction to spheroid and comparison with other works}
\label{Sec:6}

In the previous section we have presented some results for spheroids. They could be computed by assuming
$h_1$ or $h_2$ sufficiently close to 1, but in the strict limit the expressions involve singular factors.
However this singularity is only apparent. The point is that although $k_1=q_1=0$ for $h_2=1$, and $k_3=q_3=0$
for $h_1=1$, (hence all $P_k(k_s)=0$), but after we expand $P_k(k_s)$ in powers of $k_s$, substitute
(\ref{ksth}) and multiply by $M_{ij}$, some factors $(1-h_1^2)$ or $(1-h_2^2)$ cancel, leaving a well
defined limit for a spheroid.
The resulting SAM expression for $h_1=1$, $h_2=h$ is
\begin{equation} \label{genpsi}
 \Psi_3(h,\theta_3) =  \frac{8\,\left( 1 - h^2\right)  \sin^2{\theta_3} \cos{\theta_3}}{35\,\left(  1 + h^2 \right)^5}
 \left(
  2 h^4 C \cos^2{\theta_3} + S \sin^2{\theta_3} \right),
\end{equation}
where
\begin{eqnarray}
   C  &=&   \frac{ 26 + 35\, h^2}{  13 + 20\, h^2 }, \\
   S  &=&  \frac{25 + 20\, h^2 + 16\, h^4 }{ 15  + 10\, h^2 + 8\, h^4 }, \label{ourCS}
 \end{eqnarray}
and the LAM expression for $h_1=h$, $h_2=1$, is simply
\begin{equation}\label{inv}
    \Psi_1(h, \theta_1) = -h^4\,\Psi_3(h^{-1},\theta_1).
\end{equation}
The factor $h^4$ marks the difference between the LAM of the $b=c < a $ spheroid (present work)
and a prolate $a=b < c$ spheroid used by other authors.

We have found interesting to compare our solution with other published results.
Remarkably, the latter can be reduced to same general form (\ref{genpsi}) of $\Psi_3$,
differing only with the particular expressions of the coefficients $C$ and $S$.

Let us begin with \citet{EfLaz:2000}. Their solution for a rectangular prism with semi-edges $a=b$ and $c=ah$
has the form of equation (\ref{genpsi}) with
\begin{equation}\label{ELCS:0}
    C  = \frac{1323}{128}, \qquad S  = \frac{105}{16}.
\end{equation}
Obviously, the prism has a higher volume than an ellipsoid with the same $a$ and $h$.
In particular, the volume integral  $\int x^4 \mathrm{d}V$ for a spheroid is smaller
by a factor $\pi/14 \approx 0.224$. Thus, we propose to use
\begin{equation}\label{ELCS:1}
    C  = \frac{\pi}{14}\,\frac{1323}{128}, \qquad S  = \frac{\pi}{14}\, \frac{105}{16},
\end{equation}
in equation (\ref{genpsi}) to make the comparison with a spheroid more even.

\citet{MoMo:03} obtained for a spheroid
\begin{equation}\label{CSMo}
    C  = 1, \qquad S  = \frac{2}{1+\nu}.
\end{equation}
Since our solutions differ only by the choice of boundary conditions, we
adopt for comparison the values from equation (\ref{CSMo}) with $\nu=1/4$, i.e. $C=1$ and $S = 8/5$.

\begin{figure}
 \begin{center}
  \includegraphics[width=8.5cm]{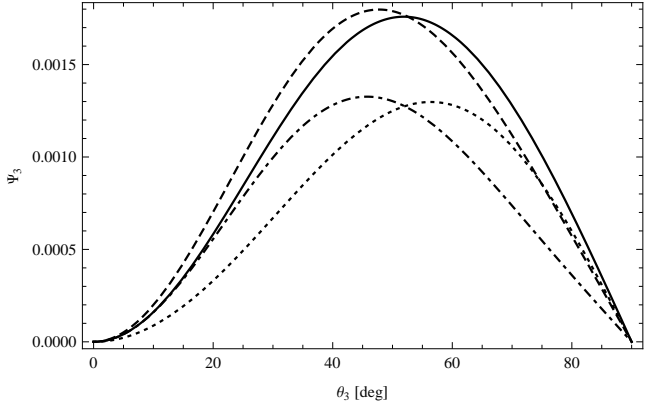}
  \end{center}
  \caption{$\Psi_3(\theta_3)$ for a spheroid with $h=c/a=0.9$.
  Solid line: present solution, dashed: volume-scaled \citet{EfLaz:2000}, dotted: \citet{MoMo:03};
  dot-dashed: re-derived \citet{SBH:05} for the equivalent $Q$ definition.}
  \label{fig:5}
\end{figure}

\begin{figure}
 \begin{center}
  \includegraphics[width=8.5cm]{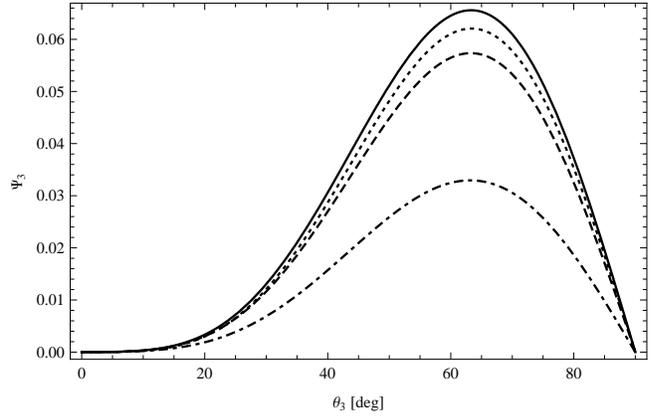}
  \end{center}
  \caption{Same as Fig.~\ref{fig:5} for $h=0.3$.}
  \label{fig:6}
\end{figure}

The complete solution of \citet{SBH:05} is known only indirectly, through the coefficients
provided by Sharma (private communication) and published in
\citet{VBNB:07}. In these circumstances, we have derived $C$ and $S$ using the stress tensor from Appendix D
of \citet{SBH:05} and following the recipe from their Sections 5 and 6  (i.e. the mainline solution, not the alternative
model). What we have obtained, agrees with
\citet{VBNB:07}, and reads
\begin{eqnarray}
  C  & = & \frac{1}{2}\, \left[  \frac{ 26 + 35\, h^2}{  13 + 20\, h^2 } \right], \nonumber \\
  S  & = & \frac{1}{4}  \, \left[ \frac{25 + 20\, h^2 + 16\, h^4 }{ 15  + 10\, h^2 + 8\, h^4 } \right], \label{sh:recon}
\end{eqnarray}
where the factors in square brackets are the same as in our present solution (\ref{ourCS}).
The difference in denominators 2 and 4 is easily understandable. First, the definition of $Q$ used by
\citet{SBH:05} is based upon the mean value of energy, hence their energy dissipation rate is twice as small
as the one based upon our more common equation~(\ref{Qdef:1}). On the other hand, \citet{SBH:05}
do not apply the multiplier $p$ in the sum given by our formula (\ref{diss:fun}), whereas $S$ is directly related
with the second harmonic of elastic energy.
In these circumstances, we will use for the comparison
\begin{equation}
  C  =  \left[  \frac{ 26 + 35\, h^2}{  13 + 20\, h^2 } \right], \quad
  S  = \frac{1}{2}  \, \left[ \frac{25 + 20\, h^2 + 16\, h^4 }{ 15  + 10\, h^2 + 8\, h^4 } \right], \label{sh:mod}
\end{equation}
i.e. the same $C$ as in our solution (\ref{ourCS}) and a half of our $S$.

Figures~\ref{fig:5} and \ref{fig:6} present $\Psi_3(\theta_3)$ of the four solutions
for oblate spheroids with $h=c/a=0.9$ and $0.3$. When the oblateness  is moderate (Fig.~\ref{fig:5}),
there is a reasonable proximity between the present model and a volume-scaled model of \citet{EfLaz:2000}, whereas
the models of \citet{MoMo:03} and \citet{SBH:05} have maxima lower by $25\%$ than the present model and shifted with respect to each other by
about $15^\circ$. Curiously, increasing the oblateness (Fig.~\ref{fig:6}), we find a good agreement in the shape of the curves
and differences of the maxima below $10\%$, with the notable exception of \citet{SBH:05} that dissipates energy
twice as slow as the remaining models. A better agreement with \citet{MoMo:03} for smaller $h$ is understandable: the only difference
with the present solution is due to different boundary conditions. They postulate a stress-free surface instead
of the usual traction-free setup, and the deviatoric part of the stress tensor on the boundary in our solution
decreases with $h$. Concerning the solution of \citet{SBH:05}, we find a systematic underestimation of $\Psi_3$ due
to the missing multiplier of the second mode, with the ratio close to 2.

Let  us remind the claims of \citet{SBH:05}, that their solution
gives damping times longer than other ones by factor 10 or more, and that the difference is due to
incomplete or incorrect solution of the elasticity problem in other papers.
Why the differences in Figs.~\ref{fig:5} and \ref{fig:6} are less drastic ?
At this point, we feel obliged to observe that for any $h$, numerical values of shape factors and damping times published and plotted
in \citet[Fig.~2]{SBH:05} differ from the ones resulting from his energy dissipation formulae by a constant factor $\pi$.
We have found no trace of this discrepancy in the published equations, so the difference, most likely,
  should be attributed to a purely computational error. Together with the incompatible definition of quality factor $Q$,
it means that all damping times from \citet{SBH:05} should be divided by $2\pi$, so they are no longer to be considered unusually high.
On the other hand, damping times shorter than Burns-Safronov estimates claimed by \citet{EfLaz:2000}
result partially (factor $14/\pi$) from using an object (a prism) with a higher volume than any
solid of revolution with the same ratio of axes.

\section{Conclusions}

Using the ensemble of standard assumptions, we have derived the stress tensor inside a freely rotating and self-gravitating ellipsoid.
Writing about known solutions to this problem, \citet{WaSch:2002} put meaningful quotation marks around the word \textit{available}.
Our present solution, given in the Appendix~\ref{ap:A} has a form which is probably compact enough  to suppress the marks,
especially for the principal axis rotation. Interestingly, the presented form of $\mtr{T}$ does not involve singularity
at the incompressible limit $\nu=\frac{1}{2}$, where \citet{WaSch:2002} had to use $\nu = 0.499$.

The stress tensor has served us as the basis for the energy dissipation model built along the lines
that \citet{Efroimsky:2002} proposed, but left unaccomplished. However,  the use of an ellipsoid
instead of an Efroimsky-Lazarian rectangular prism permitted to avoid all objections related with
partially satisfied boundary conditions and/or missing compatibility conditions. We have also found no reasons
to impose superficial conditions of a stress-free boundary like \citet{MoMo:03}.

The solution hinges upon the use of $P_k(\mathrm{q})$ series the Jacobi nome $\mathrm{q}$. Their convergence is very good
and even taking $p \leqslant 4$ in equations (\ref{P:1}-\ref{P:4}), guarantees at least three significant digits
in the area under $\Psi_s(\theta_s)$ for $0 < \theta_s < \pi/2$. Of course, the series are not legitimate
exactly at $\mathrm{q}_s=1$ (i.e $\theta_s = 90^\circ$). An in-depth discussion of this limit was given by \citet{Efroimsky:2001}.
On the other hand,  this state is not to be considered seriously, since any additional torque will
trigger the emergence of a chaotic zone in the vicinity of separatrices.

In contrast to the results of \citet{MoMo:03}, we find that the role of compressibility
in the energy dissipation process is marginal in the range of Poisson's ratio $0 \leqslant \nu \leqslant \frac{1}{4}$.
Apparently, the stronger dependence obtained by \citet{MoMo:03}
for a spheroid resulted from too strong boundary conditions.

Investigating the spheroid as a particular case of our model, we have succeeded to resolve a major part
of controversies concerning short damping times of \citet{EfLaz:2000} and very long ones according to
\citet{SBH:05}. In our opinion, the excess of energy dissipation rate over mainstream models is
mostly due to a higher volume of the body shape assumed by \citet{EfLaz:2000}. The shape factors
reported by \citet{SBH:05} are overestimated mostly by an incompatible quality factor definition,
a spurious factor $\pi$ in their computations, and a missing second mode multiplier.
We find the objections against boundary conditions used or compatibility conditions violation, raised by \citet{MoMo:03} or \citet{SBH:05},
formally justified, yet we note that they affect the accuracy of the damping/excitation times by at most $50\%$.

\section{Acknowledgements}

We thank Ishan Sharma for interesting discussions about his work and helpful comments concerning the preliminary
version of this paper.
The work of S.B. was supported by the National Science Centre grant NN 203404139. The work of D.V. was supported by grants GACR
205/08/0064 and by the Research Program MSM0021620860 of the Ministry of
Education.

Ms. Magda Murawiecka  has helped to detect the errors corrected in this version.

\bibliographystyle{mn2e}


\begin{thebibliography}{}

\bibitem[\protect\citeauthoryear{{Breiter}, {Ro{\.z}ek} \&
  {Vokrouhlick{\'y}}}{{Breiter} et~al.}{2011}]{BRV:2011}
{Breiter} S.,  {Ro{\.z}ek} A.,    {Vokrouhlick{\'y}} D.,  2011, MNRAS,
  417, 2478

\bibitem[\protect\citeauthoryear{{Burns} \& {Safronov}}{{Burns} \&
  {Safronov}}{1973}]{BuSaf:73}
{Burns} J.~A.,  {Safronov} V.~S.,  1973, MNRAS, 165, 403

\bibitem[\protect\citeauthoryear{Byrd \& Friedman}{Byrd \&
  Friedman}{1954}]{BF:1954}
Byrd P.~F.,  Friedman M.~D.,  1954, Handbook of Elliptic Integrals for
  Engineers and Scientists.
Springer, Berlin

\bibitem[\protect\citeauthoryear{Chandrasekhar}{Chandrasekhar}{1969}]{Chan:69}
Chandrasekhar S.,  1969, Ellipsoidal Figures of Equilibrium.
Yale University Press, New Haven

\bibitem[\protect\citeauthoryear{{Chree}}{{Chree}}{1895}]{Chree:1895}
{Chree} C.,  1895, Quart. J. Pure Appl. Math., 27, 338

\bibitem[\protect\citeauthoryear{{Denisov} \& {Novikov}}{{Denisov} \&
  {Novikov}}{1987}]{DeNo:87}
{Denisov} G.~G.,  {Novikov} V.~V.,  1987, Akademiia Nauk SSSR, Izvestiia,
  Mekhanika Tverdogo Tela, pp 69--74

\bibitem[\protect\citeauthoryear{{Deprit} \& {Elipe}}{{Deprit} \&
  {Elipe}}{1993}]{DepEl:93}
{Deprit} A.,  {Elipe} A.,  1993, Journal of the Astronautical Sciences, 41, 603

\bibitem[\protect\citeauthoryear{{Efroimsky}}{{Efroimsky}}{2000}]{Efroimsky:2000}
{Efroimsky} M.,  2000, J. Math. Phys., 41, 1854

\bibitem[\protect\citeauthoryear{{Efroimsky}}{{Efroimsky}}{2001}]{Efroimsky:2001}
{Efroimsky} M.,  2001, Plan. Space. Sci., 49, 937

\bibitem[\protect\citeauthoryear{{Efroimsky}}{{Efroimsky}}{2002}]{Efroimsky:2002}
{Efroimsky} M.,  2002, Advances in Space Research, 29, 725

\bibitem[\protect\citeauthoryear{{Efroimsky}}{{Efroimsky}}{2012}]{Efro:2012}
{Efroimsky} M.,  2012, Celest. Mech. and Dynamical
  Astron., 112, 283

\bibitem[\protect\citeauthoryear{{Efroimsky} \& {Lazarian}}{{Efroimsky} \&
  {Lazarian}}{2000}]{EfLaz:2000}
{Efroimsky} M.,  {Lazarian} A.,  2000, MNRAS, 311, 269

\bibitem[\protect\citeauthoryear{{Efroimsky} \& {Williams}}{{Efroimsky} \&
  {Williams}}{2009}]{EfroWil:09}
{Efroimsky} M.,  {Williams} J.~G.,  2009, Celest. Mech. and Dynamical
  Astron., 104, 257

\bibitem[\protect\citeauthoryear{{Harris}}{{Harris}}{1994}]{Harris:94}
{Harris} A.~W.,  1994, Icarus, 107, 209

\bibitem[\protect\citeauthoryear{{Innes}}{{Innes}}{1902}]{Innes:1902}
{Innes} R.~T.~A.,  1902, MNRAS, 62, 493

\bibitem[\protect\citeauthoryear{Lambeck}{Lambeck}{1980}]{Lambeck:80}
Lambeck K.,  1980, The Earth's Variable Rotation: Geophysical Causes and
  Consequences.
Cambridge University Press, Cambridge

\bibitem[\protect\citeauthoryear{Landau \& Lifshitz}{Landau \&
  Lifshitz}{1959}]{LaLi:7}
Landau L.~D.,  Lifshitz E.~M.,  1959, Theory of Elasticity.
Pergamon Press, Oxford

\bibitem[\protect\citeauthoryear{{Lazarian} \& {Efroimsky}}{{Lazarian} \&
  {Efroimsky}}{1999}]{LazEf:1999}
{Lazarian} A.,  {Efroimsky} M.,  1999, MNRAS, 303, 673

\bibitem[\protect\citeauthoryear{Love}{Love}{1934}]{Love:34}
Love A. E.~H.,  1934, A Treatise on the Mathematical Theory of Elasticity.
Cambridge University Press, Cambridge

\bibitem[\protect\citeauthoryear{{Molina}, {Moreno} \&
  {Mart{\'{\i}}nez-L{\'o}pez}}{{Molina} et~al.}{2003}]{MoMo:03}
{Molina} A.,  {Moreno} F.,    {Mart{\'{\i}}nez-L{\'o}pez} F.,  2003, A\&A, 398,
  809

\bibitem[\protect\citeauthoryear{{Munk} \& {MacDonald}}{{Munk} \&
  {MacDonald}}{1960}]{MuMa:60}
{Munk} W.~H.,  {MacDonald} G.~J.~F.,  1960, {The rotation of the Earth. A
  geophysical discussion}.
Cambridge Universtity Press, Cambridge

\bibitem[\protect\citeauthoryear{O'Connell \& Budiansky}{O'Connell \&
  Budiansky}{1978}]{OCB:78}
O'Connell R.~J.,  Budiansky B.,  1978, Geophys. Res. Let., 5, 5

\bibitem[\protect\citeauthoryear{{Paolicchi}, {Burns} \&
  {Weidenschilling}}{{Paolicchi} et~al.}{2002}]{PBW:02}
{Paolicchi} P.,  {Burns} J.~A.,    {Weidenschilling} S.~J.,  2002, Asteroids
  III, pp 517--526

\bibitem[\protect\citeauthoryear{{Pravec}, {Harris}, {Scheirich}, {Ku{\v
  s}nir{\'a}k}, {{\v S}arounov{\'a}}, {Hergenrother}, {Mottola}, {Hicks},
  {Masi}, {Krugly} \& {\& 10 coauthors}}{{Pravec} et~al.}{2005}]{Pravec:05}
{Pravec} P.,  {Harris} A.~W.,  {Scheirich} P.,  {Ku{\v s}nir{\'a}k} P.,  {{\v
  S}arounov{\'a}} L.,  {Hergenrother} C.~W.,  {Mottola} S.,  {Hicks} M.~D.,
  {Masi} G.,  {Krugly} Y.~N.,    {\& 10 coauthors} 2005, Icarus, 173, 108

\bibitem[\protect\citeauthoryear{{Prendergast}}{{Prendergast}}{1958}]{Prend:58}
{Prendergast} K.~H.,  1958, AJ, 63, 412

\bibitem[\protect\citeauthoryear{{Purcell}}{{Purcell}}{1979}]{Purc:1979}
{Purcell} E.~M.,  1979, ApJ, 231, 404

\bibitem[\protect\citeauthoryear{{Rubincam}}{{Rubincam}}{2000}]{Rub:2000}
{Rubincam} D.~P.,  2000, Icarus, 148, 2

\bibitem[\protect\citeauthoryear{Saad}{Saad}{2005}]{Saad:05}
Saad M.~H.,  2005, Elasticity. Theory, Applications and Numerics.
Elsevier, Amsterdam

\bibitem[\protect\citeauthoryear{{Scheeres}, {Ostro}, {Werner}, {Asphaug} \&
  {Hudson}}{{Scheeres} et~al.}{2000}]{SOWAH:2000}
{Scheeres} D.~J.,  {Ostro} S.~J.,  {Werner} R.~A.,  {Asphaug} E.,    {Hudson}
  R.~S.,  2000, Icarus, 147, 106

\bibitem[\protect\citeauthoryear{{Sharma}, {Burns} \& {Hui}}{{Sharma}
  et~al.}{2005}]{SBH:05}
{Sharma} I.,  {Burns} J.~A.,    {Hui} C.-Y.,  2005, MNRAS, 359, 79

\bibitem[\protect\citeauthoryear{{Tokis}}{{Tokis}}{1974}]{Tokis:74}
{Tokis} J.~N.,  1974, Ap\&SS, 26, 447

\bibitem[\protect\citeauthoryear{{Vokrouhlick{\'y}}, {Breiter}, {Nesvorn{\'y}}
  \& {Bottke}}{{Vokrouhlick{\'y}} et~al.}{2007}]{VBNB:07}
{Vokrouhlick{\'y}} D.,  {Breiter} S.,  {Nesvorn{\'y}} D.,    {Bottke} W.~F.,
  2007, Icarus, 191, 636

\bibitem[\protect\citeauthoryear{{Warner}, {Harris} \& {Pravec}}{{Warner}
  et~al.}{2009}]{Warner:09}
{Warner} B.~D.,  {Harris} A.~W.,    {Pravec} P.,  2009, Icarus, 202, 134

\bibitem[\protect\citeauthoryear{{Washabaugh} \& {Scheeres}}{{Washabaugh} \&
  {Scheeres}}{2002}]{WaSch:2002}
{Washabaugh} P.~D.,  {Scheeres} D.~J.,  2002, Icarus, 159, 314

\bibitem[\protect\citeauthoryear{Whittaker}{Whittaker}{1952}]{Whit:52}
Whittaker E.~T.,  1952, A Treatise on the Analytical Dynamics.
Cambridge Univ. Press, Cambridge

\bibitem[\protect\citeauthoryear{Wilmanski}{Wilmanski}{2010}]{Wilma}
Wilmanski K.,  2010, Fundamentals of Solid Mechanics.
IUSS Press, Pavia

\end{thebibliography}

\appendix
\section{Ellipsoid stress tensor}
\label{ap:A}

In order to shorten the expressions of the elements of $\mtr{A}$ and $\mtr{A}^{ij}$ in equation~(\ref{Tform}), we first introduce
\begin{equation}\label{Bijk}
    B^{ijk} = \frac{1}{2} \left( (-1)^i B_{11} + (-1)^j B_{22} + (-1)^k B_{33} \right),
\end{equation}
and
\begin{equation}
h_{12} = h_1 h_2.
\end{equation}
Then, we can explicitly define
\begin{eqnarray}
A^{11}_{22} &=& h_{1}^2 A_{11} + 2 A_{22} - h_{2}^{-2} A_{33} - h_{1}^2 B^{001}, \\
A^{11}_{23} &=& 3 A_{23} - h_{12}^2 B_{23},\\
A^{11}_{33} &=& h_{12}^2 A_{11} - h_{2}^2 A_{22} + 2 A_{33} - h_{12}^2 B^{010},\\
A^{22}_{11} &=& 2 A_{11} + h_1^{-2} A_{22}  - h_{12}^{-2} A_{33} - B^{001},\\
A^{22}_{13} &=& 3 A_{13} - h_{12}^2 B_{13},\\
A^{22}_{33} &=& -h_{12}^2 A_{11} + h_{2}^2 A_{22} + 2 A_{33} - h_{12}^2 B^{100},\\
A^{33}_{11} &=& 2 A_{11} - h_{1}^{-2} A_{22}  + h_{12}^{-2} A_{33} - B^{010},\\
A^{33}_{12} &=& 3 A_{12} - h_{1}^2 B_{12},\\
A^{33}_{22} &=& -h_{1}^2 A_{11} + 2 A_{22} + h_{2}^{-2} A_{33} - h_{1}^2 B^{100},\\
A^{12}_{12} &=& -h_1 \left( A_{11} + h_1^{-2} A_{22}  - h_{12}^{-2} A_{33}  -  B^{001} \right),\\
A^{12}_{13} &=& -h_1^{-1} \left( 2 A_{23} -  h_{12}^2 B_{23}\right), \\
A^{12}_{23} &=& -h_{1}\left(2 A_{13} - h_{12}^2 B_{13} \right),\\
A^{12}_{33} &=& 2 h_{12}^2 \left(2 h_1^{-1} A_{12} -  h_{1} B_{12}\right),\\
A^{23}_{11} &=& 2 h_1^{-1}\left(2 h_{12}^{-1} A_{23} -  h_{12} B_{23}\right),\\
A^{23}_{12} &=& - h_1 \left( 2 h_{12}^{-1} A_{13} -  h_{12} B_{13}\right),\\
A^{23}_{13} &=& - h_{12} \left(2 h_1^{-1} A_{12}  - h_{1}   B_{12} \right),\\
A^{23}_{23} &=& h_{1} h_{12}\left( A_{11} - h_1^{-2} A_{22} - h_{12}^{-2} A_{33}  +   B^{100} \right),\\
A^{13}_{12} &=& - 2 h_{12}^{-1} A_{23}  + h_{12} B_{23},\\
A^{13}_{13} &=& -h_{12} \left( A_{11} - h_{1}^{-2} A_{22} + h_{12}^{-2} A_{33}  - B^{010} \right),\\
A^{13}_{22} &=& 2 h_1^2 \left(2 h_{12}^{-1} A_{13} -   h_{12} B_{13} \right),\\
A^{13}_{23} &=& - h_{12} \left( 2A_{12} - h_{1}^2  B_{12} \right).
\end{eqnarray}
For the remaining 15 matrix elements that are not given above, we have (repeated index marks a pattern, no summation implied)
\begin{equation}
  A^{ii}_{jk} = A_{jk},  \quad  A^{ii}_{jj} = A_{jj}, \quad A^{ij}_{kk} =   0.
\end{equation}
The off-diagonal `central stress' elements are given directly by
\begin{eqnarray}
  A_{12} &=& \left( 1 - \frac{1+\nu}{2 h_{12}^2 h_2^2 + ( 3+  h_2^2 +h_{12}^2 )(1+\nu)  }\right) \frac{h_1^2 B_{12}}{2}, \\
  A_{13} &=& \left( 1 - \frac{(1+\nu) h_2^2 }{2 h_1^2 + ( 1 + 3 h_2^2 +h_{12}^2 )(1+\nu)  }\right) \frac{h_{12}^2 B_{13}}{2},\\
  A_{23} &=& \left( 1 - \frac{(1+\nu)h_1^2 h_{12}^2 }{2   + ( h_1^2 + h_{12}^2 + 3 h_1^2 h_{12}^2 )(1+\nu)  }\right) \frac{h_{12}^2 B_{23}}{2}.
\end{eqnarray}
The rest is obtained from equations~(\ref{ouch}) with
\begin{eqnarray}
  L_{11} &=&  - 2 - h_1^2 - h_1^4 - h_{12}^2 (1-h_1^2) \nu,\\
  L_{12} &=& -1 - h_1^2 - 2 h_1^4 + h_{12}^2 (1-  h_1^2) \nu, \\
  L_{13} &=&    1 + h_1^2 + h_1^4 + h_1^2 ( 1  + 2 h_2^2  + 2 h_{12}^2)\nu,\\
  L_{21} &=& h_1^2 (1 + h_2^2 + h_2^4) +( 2  + 2 h_2^2  + h_{12}^2 ) \nu, \\
  L_{22} &=& - h_1^2 (2 + h_2^2 + h_2^4) - (1 - h_2^2) \nu,\\
  L_{23} &=& - h_1^2 (1 + h_2^2 + 2 h_2^4) + (1 - h_2^2) \nu,\\
  L_{31} &=& -2 - h_{12}^2 - h_{12}^4 - h_1^2 (1 - h_{12}^2) \nu, \\
  L_{32} &=& 1 + h_{12}^2 + h_{12}^4 + h_1^2 ( 2 +  h_2^2 + 2 h_{12}^2) \nu,\\
  L_{33} &=&  -1 - h_{12}^2 - 2 h_{12}^4 + h_1^2 (1 - h_{12}^2) \nu,
\end{eqnarray}
and
\begin{equation}\label{Qma}
    \mtr{R} = \left(
                \begin{array}{ccc}
                  1 - h_1^2 \nu & h_1^2 \left(h_1^2 - \nu\right) & -h_{12}^2 \left(1 + h_1^2\right)  \nu \\
                  -\left(1 + h_2^2\right) \nu & h_1^2 \left(1 - h_2^2 \nu\right) & h_{12}^2 \left(h_2^2 - \nu\right) \\
                  1 - h_{12}^2 \nu & -h_1^2 \left(1 + h_{12}^2\right) \nu &  h_{12}^2 \left( h_{12}^2  -   \nu  \right) \\
                \end{array}
              \right).
\end{equation}

\section{Coefficients $M$}

\label{ap:B}
After setting the Poisson's ratio $\nu = \frac{1}{4}$, and defining
\begin{eqnarray}
  N  &=& \frac{32}{35} \left(\frac{  h_{12}^2}{ \left(1-h_1^2\right)
   \left(1-h_2^2\right)  \left(1-h_{12}^2\right),
    }\right)^2,
\end{eqnarray}
where $h_{12}=h_1h_2$, we obtain a compact form of three coefficients required in equations~(\ref{dE3}) and (\ref{dE1})
\begin{eqnarray}
M_{13} & = & N  \left(1 - h_1^4\right) \left(1 - h_2^4\right)  \left( 2 - \frac{5 h_2^2}{5 + 8 h_1^2 + 15 h_2^2 +
    5 h_{12}^2 } \right), \\
M_{23} &=& N   \left(1 - h_{12}^4\right) \left(1 - h_1^4\right)
\left( 2 - \frac{5 h_1^4 h_2^2}{8 +  5 h_1^2   +
    5 h_{12}^2 \left( 1+ 3 h_1^2\right)}\right),  \\
M_{12} &=& N   \frac{ \left(1 - h_{12}^4\right)\left(1- h_2^4\right)}{h_2^4}
\left( 2 - \frac{5}{15 +  5 h_2^2  + h_{12}^2 \left(
      5 + 8 h_2^2\right)}\right).
\end{eqnarray}
The expression of the fourth one is more involved:
\begin{eqnarray}
M_0 & = & \frac{N}{3   h_2^4 N_9} \sum_{j=0}^8 N_j\, h_2^{2j}.~~\mbox{\textbf{(corrected)}}
\end{eqnarray}
Using an auxiliary variable
\begin{equation}\label{xi}
    \xi = \left( h_1 + h_1^{-1}\right)^2,
\end{equation}
we can compress $N_j$ to read
\begin{eqnarray}
N_0 &  = & 225   \left(\xi-1\right),\\
N_1 &  = & 6   \left(1 + h_1^2\right) \left( 29 \xi-21\right),\\
N_2 &  = & h_1^2 \left(31 \xi^2 + 82 \xi -62 \right),\\
N_3 &  = & h_1^2 \left(1 + h_1^2\right) \left(- 92 \xi^2 + 305 \xi -216 \right),\\
N_4 &  = & h_1^4 \left(31 \xi^3- 341 \xi^2  + 99 \xi  + 295\right),\\
N_5 &  = & h_1^4 \left(1 + h_1^2\right) \left( 174 \xi^3  - 1012 \xi^2 + 1185 \xi -458\right),\\
N_6 &  = & h_1^6 \left( 225 \xi^4 - 1404 \xi^3+ 2412 \xi^2 - 1409 \xi  -124\right),\\
N_7 &  = & h_1^6 \left(1 + h_1^2\right) \left( 225 \xi^3 - 1179 \xi^2 + 1376 \xi -368\right),\\
N_8 &  = & h_1^8 \left( 3 \xi - 4 \right) \left( 75 \xi^2 - 292 \xi +64\right),\\
 N_9 &=&    48 \xi - 57   + h_1^2 h_2^4 \left(48 \xi^2- 119 \xi + 100   \right)
 \nonumber \\
   & &   + h_2^2 \left(1 + h_1^2\right) \left(  32 \xi-23
   +   h_1^2 h_2^4 \left(  39 \xi -44 \right) \right) \nonumber \\
  & &  + 16 h_1^4 h_2^8 ( 3 \xi-4 ).
\end{eqnarray}
\label{lastpage}
\end{document}